\newif\ifShowKeys
\ShowKeystrue
\ShowKeysfalse

\newif\ifshowtikz
\showtikztrue
\showtikzfalse   

\documentclass[11pt]{article}
	\pdfoutput=1
\usepackage[T1]{fontenc}

\usepackage{listings}
\lstnewenvironment{arkady}  
{\lstset{language=C,frame=trbl,basicstyle = \footnotesize \ttfamily , breaklines = true,showstringspaces=false}}{}

\usepackage{froufrou}
\usepackage{datetime}
\usepackage{comment}					

\usepackage{color}
\definecolor{Maroon}{rgb}{0.15,0.33,0.52}
\definecolor{Mahogany}{rgb}{0.65,0.0975,0.0845}
\usepackage[setpagesize=false,pagebackref=false, 
linktocpage, bookmarksopen=true, colorlinks=true, 
linkcolor=Maroon,citecolor=Maroon,urlcolor=Maroon]{hyperref}

\usepackage[parsep]{collref}				

\ifShowKeys \usepackage[notcite]{showkeys} \fi

\usepackage{amsmath, amssymb,amsthm}
\usepackage{stackrel}
\numberwithin{equation}{section}
\usepackage{bm,environ,mathrsfs,array,arydshln}
\usepackage{booktabs,float,slashed}
\usepackage{appendix}
\usepackage[mathcal]{euscript}
\usepackage{tensor} 						
\usepackage{mathabx}
\usepackage[vcentermath]{youngtab}
\usepackage{simpler-wick}

\usepackage{graphicx,epsfig,epic}
\usepackage{subcaption,wrapfig}
\usepackage{tikz}
\usepackage{tikz-feynman} 

\allowdisplaybreaks

\usepackage{framed}						
\definecolor{shadecolor}{rgb}{0.9996078, 0.984314, 0.960784}
\definecolor{framecolor}{rgb}{0,0,0}
\definecolor{TFTitleColor}{RGB}{1,1,1}
\definecolor{TFFrameColor}{RGB}{249	218	181}		
\definecolor{TFFrameColor}{RGB}{230 230 230 }

\newenvironment{frshaded}{%
    \MakeFramed {\FrameRestore}}%
    {\endMakeFramed}

\definecolor{myred}{RGB}{233, 33, 45}

\usepackage{mdframed}

\definecolor{lightpeach}{RGB}{255, 247, 235}

\newmdenv[
  backgroundcolor=gray!5,
  linecolor=gray!40,
  roundcorner=5pt,
  innerleftmargin=8pt,
  innerrightmargin=8pt,
  innertopmargin=6pt,
  innerbottommargin=6pt,
]{remarkbox}

\newcommand{\bs}{\begin{frshaded}}			
\newcommand{\es}{\end{frshaded}\noindent}

\def\ba#1\ea{\begin{align}#1\end{align}}		        
\newcommand{\be}{\begin{equation}}
\newcommand{\ee}{\end{equation}}
\newcommand{\bea}{\begin{equation} \begin{aligned}} 
\newcommand{\eea}{\end{aligned} \end{equation}}
\newcommand{\mc}{\mathcal }

\newcommand{\wt}{\widetilde}

\newcommand{\la}{\label}

\newcommand{\lp}{\notag \\ & }

\DeclareMathOperator{\Tr}{\text{Tr}}

\newcommand{\cf}{\textit{cf.} }
\newcommand{\ie}{\textit{i.e.} }

\newcommand{\sql}{\sqrt\l}
\renewcommand{\l}{\lambda}

\newcommand{\ket}[1]{|#1\rangle}

\newcommand{\mmm}[3]{\langle #1|#2|#3\rangle}

\newcommand{\alp}{\alpha}

\newcommand{\sD}{{\mathscr D}}

\newcommand{\qq}{{\mathsf q}}

\newcommand{\bjt}{\beta_{\rm JT}}

\makeatletter
\DeclareFontFamily{OMX}{MnSymbolE}{}
\DeclareSymbolFont{MnLargeSymbols}{OMX}{MnSymbolE}{m}{n}
\SetSymbolFont{MnLargeSymbols}{bold}{OMX}{MnSymbolE}{b}{n}
\DeclareFontShape{OMX}{MnSymbolE}{m}{n}{
    <-6>  MnSymbolE5
   <6-7>  MnSymbolE6
   <7-8>  MnSymbolE7
   <8-9>  MnSymbolE8
   <9-10> MnSymbolE9
  <10-12> MnSymbolE10
  <12->   MnSymbolE12
}{}
\DeclareFontShape{OMX}{MnSymbolE}{b}{n}{
    <-6>  MnSymbolE-Bold5
   <6-7>  MnSymbolE-Bold6
   <7-8>  MnSymbolE-Bold7
   <8-9>  MnSymbolE-Bold8
   <9-10> MnSymbolE-Bold9
  <10-12> MnSymbolE-Bold10
  <12->   MnSymbolE-Bold12
}{}

\let\llangle\@undefined
\let\rrangle\@undefined
\DeclareMathDelimiter{\llangle}{\mathopen}%
                     {MnLargeSymbols}{'164}{MnLargeSymbols}{'164}
\DeclareMathDelimiter{\rrangle}{\mathclose}%
                     {MnLargeSymbols}{'171}{MnLargeSymbols}{'171}
\makeatother

\begin{document}

\begin{titlepage}


\vspace*{15mm}
\begin{center}
{\Large Modular structures in the DSSYK partition function} 
\vspace*{10mm}

E. Alfinito$^{a,b}$ and M. Beccaria$^{a,b}$

\vspace*{4mm}
{\small
	
${}^a$ Universit\`a del Salento, Dipartimento di Matematica e Fisica \textit{Ennio De Giorgi},
\vskip 0.2cm
${}^{b}$ INFN - sezione di Lecce, Via Arnesano, I-73100 Lecce, Italy
\vskip 0.3cm
\vskip 0.2cm {\small E-mail: \texttt{matteo.beccaria@le.infn.it}}
}
\vspace*{0.8cm}
\end{center}

\begin{abstract}  
We study the low-temperature expansion of the disk partition function $Z(\beta)$ of the double-scaled SYK model 
(DSSYK) at fixed coupling $\l=2p^{2}/N$, where $N$ is the  number of Majorana fermions and $p$ is the number of fermions in each interaction term,
both taken to infinity.
We show that the exact Bessel-function representation of $Z(\beta)$, expanded at large argument (corresponding to low temperature),
can be organized  in terms of the classical ring of quasi-modular 
Eisenstein series $E_{2},E_{4},E_{6}$ and their  differential identities. Exploiting the modular $S$-duality properties of this ring, we derive the 
semiclassical (small $\l$) low-temperature expansion of $Z(\beta)$, splitting it into a perturbative tower and 
a non-perturbative sector controlled by $\wt q=e^{-4\pi^{2}/\l}$. 
At each order in $\wt q$, we determine the non-perturbative correction in closed form up to second order in $\l$;
the resulting series resums into a compact expression in the same Eisenstein series, 
extending previous semiclassical results beyond their strict $\beta\to\infty$ limit.
We further show that this entire structure follows from a single, 
exact differential equation coupling a modular derivative to derivatives with respect to temperature.
Finally, we prove that the non-perturbative sector of $Z(\beta)$  is exactly 
supported, to all orders in $\l$, on the same exponents as the on-shell actions of known bilocal-Liouville saddles 
of the DSSYK Schwarzian limit, pointing to a well-defined bulk origin for these non-perturbative corrections.
\end{abstract}
\end{titlepage}


{\small
\makeatletter
\newcommand*{\toccontents}{\@starttoc{toc}}
\makeatother
\toccontents
}


\vspace{1cm}

\setcounter{footnote}{0}

\section{Introduction and summary of results}

Since its introduction, the Sachdev-Ye-Kitaev model of $N$ strongly-coupled, randomly-interacting Majorana fermions \cite{Sachdev:1992fk,Sachdev:2010um} 
has been a rare example of a quantum-mechanical system that is simultaneously maximally chaotic and exactly solvable at large $N$ \cite{Maldacena:2016hyu}. 
Keeping the number $p$ of fermions per interaction term fixed as $N\to\infty$, the model flows in the infrared to Schwarzian quantum mechanics, the boundary dynamics of nearly-$AdS_{2}$ 
JT gravity \cite{Maldacena:2016hyu,Maldacena:2016upp,Jensen:2016pah,Sarosi:2017ykf,Berkooz:2018jqr,Mertens:2017mtv,Engelsoy:2016xyb,Mertens:2022irh}. 
Instead, a different and equally tractable regime is reached by sending $p\to\infty$ together with $N$, at fixed ratio $\l=2p^{2}/N$ \cite{Erdos:2014zgc,Cotler:2016fpe}.
In this double-scaled SYK (DSSYK) model, the disorder average can be performed exactly, reducing every observable to a combinatorial sum over intersecting 
chord diagrams \cite{Berkooz:2018jqr,Berkooz:2018qkz}, or equivalently to matrix elements of simple operators built from $q$-deformed oscillators acting on 
an auxiliary chord Hilbert space, with $q=e^{-\l}$ \cite{Lin:2022rbf,Lin:2023trc}. This exact control at all values of $\l$, well beyond the Schwarzian/JT corner 
of parameter space, is what makes DSSYK a valuable testing ground for holography. 
At generic $\l$, this holographic role has been made precise: the DSSYK transfer matrix has been matched to a canonical transformation of the bulk Hamiltonian 
of two-dimensional sine-dilaton gravity, identifying the latter as the exact holographic dual of the model 
\cite{Blommaert:2023opb,Blommaert:2024ymv,Blommaert:2024whf,Heller:2024ldz,Heller:2025ddj}, as checked explicitly at 1-loop
in \cite{Bossi:2024ffa} and at higher-loop order and infinite temperature in \cite{Alfinito:2026cky}, see also \cite{Fu:2025kkh}.

In this paper we reconsider the disk partition function $Z(\beta)$, and specifically  its behaviour at low temperature, $\beta\to\infty$, at fixed double-scaling parameter $\l$. 
The high-temperature ($\beta\to0$) expansion of $Z(\beta)$ is a convergent power series in $\beta$ with $\l$-dependent coefficients, reflecting the
analyticity of the exact partition function at $\beta=0$ \cite{Okuyama:2023iwu,Verg2013}. The low-temperature regime is comparatively less studied and qualitatively different: 
the semiclassical ($\l\to0$) saddle-point expansion of $Z(\beta)$ at generic $\beta$ \cite{Goel:2023svz,Okuyama:2023bch} is only asymptotic in $\l$, 
and matching it order by order onto the small-$\beta$ result shows that a genuine non-perturbative completion, invisible to this saddle-point expansion, is required once $\beta$ becomes large.

An initial treatment of the non-perturbative low-temperature completion was given in \cite{Okuyama:2025fhi}. 
Starting from the exact Bessel-function representation of $Z(\beta)$ \cite{Berkooz:2018jqr,Xu:2024gfm}, 
and using the leading large-argument behaviour of the Bessel function, it was shown that the semiclassical (Schwarzian-type) 
partition function is exactly multiplied, to all orders in $\wt q=e^{-4\pi^{2}/\l}$, by the cube of the Dedekind eta function 
$\eta(\wt q)^{3}$, with no residual dependence on $\beta$ at the order of approximation considered there. 
A possible holographic interpretation of this correction was also proposed there.

The present paper extends this result and places it inside a more complete, largely modular structure, with three main results. 
The first is that the Bessel-function representation of $Z(\beta)$, expanded at large values of its natural argument $x=2\beta/\sqrt{\l(1-q)}$, can be organized 
entirely in terms of the classical ring of quasi-modular Eisenstein series $E_{2},E_{4},E_{6}$ and their Ramanujan differential identities, eq.~(\ref{4.22}). 
Exploiting the modular ($S$-duality) inversion properties of $E_{2n}$, we re-expand this result in the double-scaled regime of small $\l$ at fixed $\beta$, 
splitting $\log Z(\beta)$ into a perturbative tower, systematic to all orders in both $1/\beta$ and $\l$, and a non-perturbative sector controlled 
by the same parameter $\wt q$ as in \cite{Okuyama:2025fhi}. 

A second, and perhaps more surprising, result follows from systematically keeping the subleading large-$\beta$ corrections neglected by the 
leading Bessel approximation of \cite{Okuyama:2025fhi}: at each order $\wt q^{n}$ the non-perturbative correction is 
controlled by a genuine function of $y=24\pi^{2}/(\l\beta)$ order by order in $\l$.
Explicitly,
\be
\la{1.1}
[\log Z(\beta)]_{\rm NP} = \sum_{n=1}^{\infty} L_{n}(\beta)\,\wt q^{n}, \qquad L_{n}(\beta) = \sum_{p=0}^{\infty} f_{n,p}(y)\,\l^{p}.
\ee
We determine these functions in closed form, 
for general $n$, up to second order in $\l$, eq.~(\ref{5.35}), with coefficients built from the divisor functions
$\sigma_{1}(n)$ and, starting at order $\l^{2}$, $\sigma_{3}(n)$. \footnote{The divisor function $\sigma_{p}(n)$ is the sum of the $p$-th powers
of all divisors of $n$.}
In particular, in the strict $\beta\to\infty$ limit, $f_{n,0}(0)=-3\sigma_{1}(n)/n$ reproduces the $\beta$-independent result 
of \cite{Okuyama:2025fhi}. Using the generating functions of $\sigma_{1}(n)$ and $\sigma_{3}(n)$ in terms of $E_{2}$ and $E_{4}$, 
the entire non-perturbative series up to $O(\l^{2})$ resums into a closed expression in these quasi-modular Eisenstein series evaluated at $\Lambda=e^{y/3}\wt q$.
Including the first $O(\l)$ correction, our result reads
\ba
\la{1.2}
[\log Z(\beta)]_{\rm NP} &= 3\,\log(\Lambda; \Lambda)_{\infty}+\frac{1-E_{2}(\Lambda)}{96}y\,\bigg(1+\frac{y}{3\pi^{2}}\bigg)\, \l +\cdots,
\ea
where the further $O(\l^{2})$ correction is given explicitly in (\ref{5.38}) and involves both $E_{2}^{2}$ and $E_{4}$. The expansion of (\ref{1.2})
in powers of $\wt q$ generates the full set of non-perturbative corrections.

A final result is that this entire structure -- perturbative and non-perturbative alike -- follows from a single, exact differential equation obeyed by the 
Bessel-function representation of the partition function, relating the modular derivative $\sD=q\,d/dq$ and $x$-derivatives, eq.~(\ref{6.3}). 
This differential equation is exact in $\l$ and $\beta$, requiring no semiclassical expansion for its derivation. 
Both the perturbative tower and the exponential, $y$-dependent structure of the non-perturbative functions $f_{n,p}(y)$ are recovered from it as solutions in the appropriate limits, 
with the divisor-sum coefficients fixed as boundary data by matching onto the quasi-modular expansion.

Finally, we show that the non-perturbative sector of $Z(\beta)$ itself -- as opposed to $[\log Z(\beta)]_{\rm NP}$, which 
receives contributions at every order in $\wt q$ -- is exactly supported, to all orders in $\l$, on a sparse set of 
exponents dictated by the ring structure of Section~\ref{sec:modular-structure}. These exponents are the triangular numbers $n=j(j+1)/2$, with $j=0,1,2,\dots$, and match precisely the 
on-shell actions of the additional saddles of the bilocal-Liouville description of the DSSYK Schwarzian limit recently 
identified in \cite{Berkooz:2024ifu}, pointing to a well-defined bulk origin for the non-perturbative corrections found here.

The rest of the paper is organized as follows. In Section~\ref{sec:partition} we review the DSSYK model and the 
exact representations of its partition function. Section~\ref{sec:saddle} discusses the semiclassical saddle-point 
expansion at generic $\beta$ and its convergent high-temperature limit. Section~\ref{sec:modular-structure} 
derives the quasi-modular large-$x$ expansion. Section~\ref{sec:semiclassical} 
performs the joint small-$\l$, large-$\beta$ expansion, separating perturbative and non-perturbative 
contributions and comparing with \cite{Okuyama:2025fhi,Berkooz:2024ifu}. Section~\ref{sec:modular} derives the 
differential equation (\ref{6.3}) and shows how it reproduces all the previous results. 
Section~\ref{sec:conclusions} summarizes our results and compares the non-perturbative 
corrections with the saddle spectrum of \cite{Berkooz:2024ifu}.
Appendix~\ref{app:resum} contains the derivation of the non-perturbative resummation.

\section{The DSSYK model and partition function}
\la{sec:partition}

The SYK model \cite{Sachdev:1992fk,Sachdev:2010um,Maldacena:2016hyu} is built from $N$ Majorana fermions $\psi_{i}$, satisfying $\{\psi_{i}, \psi_{j}\}=2\delta_{ij}$, $i,j=1, \dots, N$,
with Hamiltonian
\bea
H_{\rm SYK} &= i^{p/2}\sum_{1\le i_{1}<i_{2}<\cdots<i_{p}\le N} J_{i_{1}\cdots i_{p}}\ \psi_{i_{1}}\cdots \psi_{i_{p}},\\
\llangle J_{i_{1}\cdots i_{p}}\rrangle_{J}&=0, \qquad
\llangle J_{i_{1}\cdots i_{p}}^{2}\rrangle_{J} =\frac{N}{2p^{2}}\binom{N}{p}^{-1}J^{2},
\eea
where $J_{i_{1}\cdots i_{p}}$ are Gaussian random couplings and double angular brackets denote ensemble averaging. Our normalization follows \cite{Maldacena:2016hyu,Lin:2022rbf,Lin:2023trc}.
The double-scaling (DS) limit \cite{Erdos:2014zgc,Cotler:2016fpe} is defined by
\be
\la{2.2}
N,p\to\infty, \qquad \l\equiv 2p^{2}/N=\text{fixed}, \qquad q\equiv e^{-\l}.
\ee
In this limit, DSSYK admits an effective Hilbert space \cite{Berkooz:2018jqr,Lin:2022rbf}, spanned by 
orthonormal chord states $\{\ket{\ell}\}_{\ell=0,1,2,\dots}$. In this basis, one introduces the Hamiltonian
\be
\la{2.3}
H = \frac{J}{\sql}(\alp+\bar\alp), 
\ee
where $\alp$ and $\bar\alp$ are $q$-deformed oscillators. \footnote{With the chord number operator $\hat{\ell}$, defined by $\hat\ell\ket{\ell} = \ell\ket{\ell}$, 
they obey \cite{Lin:2023trc}
$[\alpha, \bar\alpha]_{q} = \alpha\bar\alpha-q\,\bar\alpha\alpha=1$, $[\hat\ell, \bar\alpha]=\bar\alpha$, and $[\hat\ell, \alpha] = -\alpha$.
}
and one proves that, in the double-scaling limit, ensemble averaging is equivalent to replacing $H_{\rm SYK}\to H$ and taking the expectation value in the 0-chord state
\be
\llangle\Tr f(H_{\rm SYK})\rrangle \stackrel{\rm DS}{\Rightarrow} \mmm{0}{f(H)}{0}.
\ee
In the holographic correspondence, the 0-chord state $\ket{0}$ is dual to the infinite-temperature thermofield-double state in the gravitational theory
\cite{Lin:2022rbf}.
The Euclidean partition function in chord space admits the spectral representation \cite{Berkooz:2018jqr,Berkooz:2018qkz}
\be
\la{2.5}
Z(\beta) = \mmm{0}{e^{-\beta H}}{0} = (q;q)_{\infty}\int_{0}^{\pi}\frac{d\theta}{2\pi} e^{-\beta\frac{2J\cos\theta}{\sqrt{\l(1-q)}}}|(e^{2i\theta}; q)_{\infty}|^{2},
\ee
where $(z;q)_{\infty} = \prod_{k=0}^{\infty}(1-z q^{k})$, and we omit the dependence on $\l$. In the following, we will set $J=1$, since the dependence on $J$ may always be reconstructed by $\beta\to J\beta$.
Another exact representation of the partition function, which will be important below, is 
the Bessel expansion \cite{Berkooz:2018jqr,Xu:2024gfm},
\ba
\la{2.6}
Z(\beta) &= F(x(\beta)), \qquad x(\beta) = \frac{2\beta}{\sqrt{\l(1-q)}},  \\
\la{2.7}
F(x) &= \sum_{m=-\infty}^{\infty}(-1)^{m}q^{\binom{m}{2}}I_{2m}(x) = \frac{2}{x}\sum_{n=0}^{\infty}(-1)^{n}(2n+1)I_{2n+1}(x)\,q^{n(n+1)/2},
\ea
where $I_{n}(x)$ is the modified Bessel function of the first kind. Note that the peculiar exponents of $q$ are the so-called triangular integers.

\section{Perturbative semiclassical saddle-point expansion at generic $\beta$}
\la{sec:saddle}

The semiclassical expansion of the partition function at fixed $\beta$ and $\l\to 0$ 
has been treated in \cite{Goel:2023svz,Okuyama:2023bch}. Up to non-perturbative corrections, the spectral representation (\ref{2.5}) can be written as \footnote{The sign of the $\beta$ term
is irrelevant since it is flipped by $\theta\to \pi-\theta$.}
\be
Z = \int_{0}^{\pi}\frac{d\theta}{2\pi}\sqrt\frac{2\pi}{\l}\exp\bigg[\frac{2\beta\cos\theta}{\sqrt{\l(1-q)}}+\frac{\l}{8}-\frac{2}{\l}\bigg(\theta-\frac{\pi}{2}\bigg)^{2}+\log(2\sin\theta)\bigg].
\ee
Expanding in small $\l$ and changing variable to $u=\frac{\pi}{2}-\theta$, the exponent in square brackets reads
\be
-\frac{2}{\l}(u^{2}-\beta\sin u)+\log(2\cos u)+\frac{1}{2}\beta \sin u+O(\l).
\ee
Differentiating with respect to $u$ gives the leading-order saddle-point equation
\be
\la{3.3}
\beta=\frac{2u}{\cos u}.
\ee
Integrating over fluctuations around the saddle gives the following expansion for $\log Z(\beta)$
\be
\la{3.4}
\log Z(\beta) = \frac{1}{\l}\sigma_{0}(u)+\sigma_{1}(u)+\sigma_{2}(u)\,\l + \cdots,
\ee
where the functions $\sigma_{n}(u)$ can be systematically computed. The first two read
\be
\la{3.5}
\sigma_{0}(u) = -2u^{2}+4u\tan u, \qquad \sigma_{1}(u) = u \tan u+\log\cos u-\frac{1}{2}\log(1+u \tan u).
\ee
This approach captures only perturbative corrections in small $\l$; any exponentially small contributions remain invisible.

\subsection{Comparison with the high-temperature expansion}

It is interesting to compare the semiclassical expansion (\ref{3.4}) with the high-temperature ($\beta\to 0$) expansion of the partition function
\cite{Okuyama:2023iwu}. To this end, it is convenient to start from the Bessel expansion (\ref{2.7}) and use the convergent small-$x$ expansion
\be
I_{2n+1}(x) = \sum_{k=0}^{\infty}\frac{1}{k!(k+2n+1)!}(x/2)^{2k+2n+1}.
\ee
We get the explicit expansion of $F(x)$ in small $x$
\ba
F(x) &= \sum_{p=0}^{\infty}(x/2)^{2p}\sum_{n=0}^{p}(-1)^{n}\frac{2n+1}{(p-n)!(p+n+1)!}\,q^{n(n+1)/2},
\ea
where each power of $x$ has a coefficient which is a polynomial in $q$. 
The first terms are
\be
F(x) = 1+\frac{1-q}{8}x^{2}+\frac{2-3q+q^{3}}{384}x^{4}+\frac{5-9q+5q^{3}-q^{6}}{46080}x^{6}+\cdots.
\ee
This series has a finite radius of convergence for $0\le q<1$ as discussed in \cite{Okuyama:2023iwu}, with no room for non-perturbative corrections of any kind,
as a consequence of the analyticity of $Z(\beta)$ at $\beta=0$.
The logarithm of the partition function takes the remarkable form 
\be
\la{3.9}
\log F(x) = \frac{1-q}{8}x^{2}-\frac{(1-q)^{3}}{384}x^{4}+\frac{(1-q)^{5}(5+q)}{46080}x^{6}-\frac{(1-q)^{7}(56+28q+7q^{2}+q^{3})}{10321920}x^{8}+\cdots,
\ee
where the structure of $(1-q)^{n-1}x^{n}$ factors is not accidental and persists at all orders, as discussed in \cite{Okuyama:2023iwu} based on \cite{Verg2013}. 
In fact, in Section~\ref{sec:high-T-from-ODE}, we will prove an explicit recursion for the coefficients in  (\ref{3.9}), \cf (\ref{6.23}).

In particular, this implies that, in the semiclassical limit $\l\to 0$ (after replacing $x\to x(\beta)$ as in (\ref{2.6})),
one may write a double expansion of the form 
\be
\log Z(\beta) = \frac{1}{\l}\,\bigg(\frac{\beta^{2}}{2}-\frac{\beta^{4}}{24}+\frac{\beta^{6}}{120}+\cdots\bigg)
+\frac{\beta^{4}}{48}-\frac{7 \beta^{6}}{720}+\frac{61 \beta^{8}}{13440}+\cdots+\bigg(-\frac{\beta ^4}{144}+\frac{\beta ^6}{144}-\frac{\beta^8}{8064}+\cdots\bigg)\, \l+O(\l^{2}).
\ee
This is in full agreement with (\ref{3.4}), via (\ref{3.5}) and the small-$\beta$ solution of (\ref{3.3})
\be
u = \frac{\beta}{2}-\frac{\beta^{3}}{16}+\frac{13 \beta^{5}}{768}+\cdots,
\ee
and shows that there are no order-of-limits problems when $\l\to 0$ and $\beta\to 0$.
Notice that we started from (\ref{2.7}), which is exact and has no non-perturbative correction in either $\beta$ or $\l$.
Non-perturbative corrections to the saddle-point calculation are negligible in this high-temperature semiclassical limit. 

At low temperature, the situation is more involved as we now discuss. The large-$\beta$ expansion 
is asymptotic with vanishing radius of convergence at any $\l$ and genuinely requires a non-trivial non-perturbative completion.

\section{Quasi-modular expansion at low temperature}
\la{sec:modular-structure}

In this section we discuss the low-temperature expansion of the partition function at fixed $\l$. To this end, we will need to 
recall some facts about the ring of quasi-modular Eisenstein series.

\subsection{Eisenstein series}

Besides the basic $q$-series
\be
(q;q)_{\infty} = \prod_{n=1}^{\infty}(1-q^{n}),
\ee 
let us also introduce the Eisenstein series defined by the following convergent expansion for $|q|<1$
\be
\la{4.2}
E_{2m}=1-\frac{4m}{B_{2m}}\sum_{n=1}^{\infty}\frac{n^{2m-1}q^{n}}{1-q^{n}}, \qquad m=1,2,\dots,
\ee
where $B_{2m}$ are the Bernoulli numbers.
We will be interested in the first three cases $m=1,2,3$, \ie
\be
E_{2}=1-24\sum_{n=1}^{\infty}\frac{n\,q^{n}}{1-q^{n}}, \qquad
E_{4}=1+240\sum_{n=1}^{\infty}\frac{n^{3}\,q^{n}}{1-q^{n}}, \qquad
E_{6}=1-504\sum_{n=1}^{\infty}\frac{n^{5}\,q^{n}}{1-q^{n}}.
\ee
These three functions, together with $(q;q)_{\infty}$, form a closed ring under the differential operator
\be
\sD = q\frac{d}{dq}, 
\ee
as follows from the Ramanujan identities
\be
\la{4.5}
\sD E_{2}=\frac{1}{12}(E_{2}^{2}-E_{4}), \qquad \sD E_{4}=\frac{1}{3}(E_{2}E_{4}-E_{6}), \qquad \sD E_{6} = \frac{1}{2}(E_{2}E_{6}-E_{4}^{2}),
\ee
and also 
\be
\la{4.6}
\sD(q;q)_{\infty} = \frac{1}{24}(E_{2}-1)\, (q;q)_{\infty}.
\ee
The above $q$-series can be expressed in terms of elliptic functions. We parametrize $q=q(\tau)$ and also introduce its square root, the elliptic nome $\qq$,
\be
\la{4.7}
q=e^{2\pi i \tau}, \qquad \mathsf \qq = q^{1/2}.
\ee
Then, we have the following relations in terms of the Dedekind function $\eta(\tau)$ and Jacobi theta functions
\bea
(q;q)_{\infty} &= q^{-1/24}\eta(\tau), \qquad\qquad
E_{2}(q) = \frac{12}{i\pi}\frac{d}{d\tau}\log\eta(\tau), \\
E_{4}(q) &= \frac{1}{2}(\vartheta_{2}^{8}+\vartheta_{3}^{8}+\vartheta_{4}^{8}),\qquad 
E_{6}(q) = -\frac{1}{2}(\vartheta_{2}^{4}+\vartheta_{3}^{4})(\vartheta_{2}^{4}-\vartheta_{4}^{4})(\vartheta_{3}^{4}+\vartheta_{4}^{4}),
\eea
where $\vartheta_{n}\equiv \vartheta_{n}(\qq) = \vartheta_{n}(q^{1/2})$ with 
\be
\la{4.9}
\vartheta_{2}(\qq) = 2\qq^{1/4}\sum_{n=0}^{\infty}\qq^{n(n+1)}, \qquad 
\vartheta_{3}(\qq) = \sum_{n\in\mathbb Z}\qq^{n^{2}}, \qquad
\vartheta_{4}(\qq) = \sum_{n\in\mathbb Z}(-1)^{n}\qq^{n^{2}}.
\ee

\paragraph{Quasi-modular transformation}

Under the $SL(2, \mathbb Z)$ transformation
\be
\tau\to \tau' = \frac{a\tau+b}{c\tau+d}, \qquad \begin{pmatrix} a & b \\ c & d \end{pmatrix}\in SL(2,\mathbb Z),
\ee
the Eisenstein series obey
\ba
\la{4.11}
E_{2}(\tau') &= (c\tau+d)^{2}E_{2}(\tau)+\frac{1}{6\pi i}c(c\tau+d), \\
E_{4}(\tau') &= (c\tau+d)^{4}E_{4}(\tau), \\
E_{6}(\tau') &= (c\tau+d)^{6}E_{6}(\tau),
\ea
where the notation is $E_{n}(\tau)\equiv E_{n}(e^{2\pi i \tau})$. The anomaly term in (\ref{4.11}) shows that $E_{2}$ is quasi-modular, while $E_{4}$ and $E_{6}$
are genuine modular forms. In particular, for the inversion transformation $\tau\to -1/\tau$, one has
\ba
\la{4.14}
E_{2}(-1/\tau) &= \tau^{2}E_{2}(\tau)-\frac{6i\tau}{\pi}, \qquad
E_{4}(-1/\tau) = \tau^{4}E_{4}(\tau), \qquad
E_{6}(-1/\tau) = \tau^{6}E_{6}(\tau).
\ea
We will also need the inversion rule for $\eta(\tau)$:
\be
\la{4.15}
\eta(-1/\tau) = \sqrt{-i\tau}\, \eta(\tau).
\ee

\subsection{Low-temperature expansion of the partition function at fixed $\l$}

Let us expand the modified Bessel functions in (\ref{2.7}) at large (real) argument $x\to +\infty$ using 
\ba
I_{2n+1}(x) &=  \sqrt\frac{1}{2\pi x}e^{x}\sum_{k=0}^{\infty}\frac{(-1)^{k}}{k!8^{k}x^{k}}\prod_{j=1}^{k}(4n+2j+1)(4n-2j+3)\lp
=\sqrt\frac{1}{2\pi x}e^{x}\bigg[1-\frac{(4n+3)(4n+1)}{8x}+\frac{(4n+5)(4n+3)(4n+1)(4n-1)}{128x^{2}}+\cdots\bigg].
\ea
Then, the partition function (in terms of the $x$ variable introduced in (\ref{2.6})) admits the expansion
\be
F(x) = \sqrt\frac{2}{\pi}\, x^{-3/2}e^{x}\sum_{n=0}^{\infty}(-1)^{n}(2n+1)\,q^{n(n+1)/2}\sum_{k=0}^{\infty}\frac{(-1)^{k}}{k!8^{k}x^{k}}\prod_{j=1}^{k}(4n+2j+1)(4n-2j+3).
\ee
The term $k=0$ requires computing the infinite sum 
\be
S_{0}(q) = \sum_{n=0}^{\infty}(-1)^{n}(2n+1)\,q^{n(n+1)/2} = (q;q)_{\infty}^{3} = q^{-1/8}\eta(\tau)^{3},
\ee
which can be verified by using the  Jacobi Triple Product Identity.
In the other terms with $k>0$, we use 
\be
\prod_{j=1}^{k}(4n+2j+1)(4n-2j+3) = \prod_{j=1}^{k}\bigg[32 \frac{n(n+1)}{2}+3-4j(j-1)\bigg],
\ee
and obtain 
\be
\la{4.20}
F(x) = \sqrt\frac{2}{\pi}x^{-3/2}e^{x}\sum_{k=0}^{\infty}\frac{(-1)^{k}}{k!8^{k}x^{k}}\prod_{j=1}^{k}[32 \sD+3-4j(j-1)]\,S_{0} .
\ee
Each term can be expressed as a polynomial in $E_{2}, E_{4}, E_{6}$, multiplied by $S_{0}$, using (\ref{4.5})
and the relation (\ref{4.6}) written in the form 
\be
\sD S_{0} = \frac{1}{8}(E_{2}-1)\, S_{0}.
\ee
The first terms of (\ref{4.20}) are 
\ba
\la{4.22}
F(x) &= \sqrt\frac{2}{\pi}x^{-3/2}e^{x}\bigg[1+\frac{1-4E_{2}}{8x}+\frac{27-120E_{2}+80E_{2}^{2}-32E_{4}}{384x^{2}}\lp
+\frac{1}{27648 x^3}(2025-9324 E_2 +8400 E_2^2-3360 E_4
-2240 E_2^3+2688 E_2 E_4-1024 E_6)\lp
+\frac{1}{294912 x^{4}}(
33075-154992 E_2 +157920 E_2^2-63168 E_4
-62720 E_2^3+75264 E_2 E_4 \lp
-28672 E_6
+8960 E_2^4-21504 E_2^2 E_4-3072 
E_4^2+16384 E_2 E_6)\lp
+\cdots\bigg]\,(q;q)_{\infty}^{3},
\ea
and we remark that this expansion is exact in $q=e^{-\l}$. \footnote{Similar structures are observed in other context, see in particular \cite{Beccaria:2024szi} and 
computations in deformed $\mc N=2$ superconformal theories \cite{Billo:2013jba,Bajc:2025jjv}.} 

The coefficients above are not homogeneous polynomials in $E_{2},E_{4},E_{6}$ -- e.g.\ the $k=1$ term mixes a constant with $E_{2}$ -- but they do obey a clean filtration. 
Let us assign the reduced weights 1, 2, 3 to $E_{2},E_{4},E_{6}$ 
(half the usual modular weight). From the ring relations (\ref{4.5}), every term on the right-hand side of $\sD E_{2m}$ has reduced weight exactly one more than $E_{2m}$ itself:
\be
\sD E_{2}\sim E_{2}^{2}-E_{4}, \quad
\sD E_{4}\sim E_{2}E_{4}-E_{6}, \quad
\sD E_{6}\sim E_{2}E_{6}-E_{4}^{2},
\ee
so $\sD$ raises reduced weight by exactly $1$, with no exceptions -- even though $E_{2}$ itself is only quasi-modular. In $[32\sD+3-4j(j-1)]$, the additive constant $3-4j(j-1)$ has reduced weight $0$: 
it does not raise the weight; it only feeds a lower-weight piece back in at each step. 
Consequently, after $k$ applications to $S_{0}$, the resulting polynomial in $E_{2},E_{4},E_{6}$ 
has reduced weight \emph{at most} $k$ (not exactly $k$) -- 
which is exactly why the constant and $E_{2}$ can coexist at $k=1$.

The space of polynomials $E_{2}^{a}E_{4}^{b}E_{6}^{c}$ of reduced weight $a+2b+3c=n$ has dimension $p_{3}(n)$, 
the number of partitions of $n$ into parts $\leq3$, with the classical closed form $p_{3}(n)=\big\lfloor\frac{(n+3)^{2}}{12}\big\rceil$ (nearest integer). The number of independent terms allowed at order $x^{-k}$ is then the cumulative count
\be
\la{4.24}
d(k) = \sum_{n=0}^{k}p_{3}(n): \qquad d(0),d(1),d(2),\dots = 1,\,2,\,4,\,7,\,11,\,16,\,23,\dots\ .
\ee
This matches, term for term, the number of monomials actually appearing in the exact coefficients above, and predicts $16$ independent terms at $k=5$ before any computation is done. 
\footnote{This bound may be  useful in practice: rather than repeating the operator algebra at each order, 
one can posit the general weight-$\le k$ ansatz with $d(k)$ unknown coefficients and fix them by 
matching either the recursion directly or a few low Bessel/$q$-expansion coefficients.}

\paragraph{Remark}

The series (\ref{4.22}) is asymptotic with zero radius of convergence since it is based on the large-argument expansion of the modified
Bessel function. Since the Borel transform of the asymptotic series for $I_{2n+1}(x)$ with $x\to +\infty$ has a singularity on the real axis, the function should be 
reconstructed by averaging the lateral Borel transforms,
leading to a correction $\sim \exp(-2x)$. This refinement is not studied further here, but
it is worth mentioning as the natural next step if one ever wants the non-perturbative completion of the large-$x$ expansion \emph{at fixed $\l$} explicitly, as opposed to 
non-perturbative corrections in small $\l$ to be discussed later.

\section{Low-temperature semiclassical expansion}
\la{sec:semiclassical}

The expansion (\ref{4.22}) can be conveniently studied in the semiclassical limit $\l\to 0$ by exploiting the 
modular inversion rules in (\ref{4.14},\ref{4.15}). For $\l>0$, we set 
\be
\tau = \frac{2\pi i}{\l}, \qquad -\frac{1}{\tau} = \frac{i\l}{2\pi},
\ee
and consider
\be
\wt q\equiv q(\tau)  = e^{-\frac{4\pi^{2}}{\l}}, \qquad q\equiv q(-1/\tau) = e^{-\l}.
\ee
The inversion relations (\ref{4.14}) then read
\ba
\la{5.3}
E_{2}(q) &= -\frac{4\pi^{2}}{\l^{2}}E_{2}(\wt q)+\frac{12}{\l}, \qquad
E_{4}(q) = \bigg(\frac{2\pi}{\l}\bigg)^{4}E_{4}(\wt q), \qquad
E_{6}(q) = -\bigg(\frac{2\pi}{\l}\bigg)^{6}E_{6}(\wt q).
\ea
Besides, we may use (\ref{4.15}) to write
\ba
\la{5.4}
(q;q)_{\infty}^{3} &=q^{-1/8}\eta(-1/\tau)^{3} =  q^{-1/8}(-i\tau)^{3/2}\eta(\tau)^{3} = (q/\wt q)^{-1/8}\bigg(\frac{2\pi}{\l}\bigg)^{3/2}(\wt q; \wt q)_{\infty}^{3}\lp
= e^{\frac{\l}{8}-\frac{\pi^{2}}{2\l}}\bigg(\frac{2\pi}{\l}\bigg)^{3/2}(\wt q; \wt q)_{\infty}^{3}.
\ea
The parameter $\wt q$ is the natural parameter for the non-perturbative corrections, which can be controlled order by order in $\wt q$ by 
expanding $E_{2n}(\wt q)$ and $(\wt q; \wt q)_{\infty}^{3}$ in powers of $\wt q$. To examine the resulting structure of the partition function, 
we begin with the perturbative part where $\wt q$ is set to zero, and later move on to the analysis of non-perturbative corrections.
The present analysis, explicit and based on the expansion (\ref{4.22}), represents a refinement of the leading-order analysis in \cite{Okuyama:2025fhi}.
The structure that will emerge will later have a deeper interpretation, 
on the basis of a novel differential equation satisfied by the partition function.

\subsection{Perturbative part}

The perturbative part is obtained by dropping any term depending on $\wt q$ which is exponentially small in the $\l\to 0$ limit.
Substituting $x\to x(\beta)$ as in (\ref{2.6}), we get 
\ba
Z(\beta) &= \sqrt\frac{2}{\pi}x^{-3/2}e^{x}e^{\frac{\l}{8}-\frac{\pi^{2}}{2\l}}\bigg(\frac{2\pi}{\l}\bigg)^{3/2}\bigg[1+\frac{c_{1}(\l)}{\beta}+\frac{c_{2}(\l)}{\beta^{2}}+\cdots\bigg]\lp
= \frac{\pi\sqrt 2\, (1-e^{-\l})^{3/4}}{\beta^{3/2}\l^{3/4}}\exp\bigg[
\frac{2\beta}{\sqrt{\l(1-e^{-\l})}}+\frac{\l}{8}-\frac{\pi^{2}}{2\l}
\bigg]\bigg[1+\frac{c_{1}(\l)}{\beta}+\frac{c_{2}(\l)}{\beta^{2}}+\cdots\bigg],
\ea
where the exact-in-$\l$ expressions of the functions $c_{n}(\l)$ are 
\bea
c_{1}(\l) &=\bigg( \frac{2\pi^{2}}{\l^{2}}-\frac{6}{\l}+\frac{1}{8} \bigg)\frac{1}{2}\sqrt{\l(1-e^{-\l})}, \\
c_{2}(\l) &=\bigg( \frac{2\pi^{4}}{\l^{4}}-\frac{20\pi^{2}}{\l^{3}}+\frac{5(24+\pi^{2})}{4\l^{2}}-\frac{15}{4\l}+\frac{9}{128} \bigg)\frac{1}{4}\l(1-e^{-\l}), 
\eea
and so on. For the logarithm of the partition function, we thus get 
\ba
\la{5.7}
\log Z(\beta) &= \log\frac{\pi\sqrt 2\, (1-e^{-\l})^{3/4}}{\l^{3/4}}-\frac{3}{2}\log\beta +
\frac{2\beta}{\sqrt{\l(1-e^{-\l})}}+\frac{\l}{8}-\frac{\pi^{2}}{2\l}\lp
+\frac{b_{1}(\beta)}{\l}+b_{2}(\beta)+b_{3}(\beta)\,\l+\cdots .
\ea
We note the non-trivial fact that the leading singularity of $\log Z$ for $\l\to 0$ is just $1/\l$, consistent with what happens 
at generic $\beta$ in (\ref{3.4}).
The expansion of the energy term in small $\l$ reads
\be
\frac{2\beta}{\sqrt{\l(1-e^{-\l})}} = \frac{2\beta}{\l}+\frac{\beta}{2}+\frac{\beta}{48}\l+\cdots,
\ee
and can be combined with the $b_{n}$ functions to give the more compact expression
\ba
\la{5.9}
\log Z(\beta) &= \log\frac{\pi\sqrt 2\, (1-e^{-\l})^{3/4}}{\l^{3/4}}-\frac{3}{2}\log\beta 
+\frac{\overline b_{1}(\beta)}{\l}+\overline b_{2}(\beta)+\overline b_{3}(\beta)\,\l+\cdots,
\ea
where
\be
\overline b_{1}(\beta) = b_{1}(\beta)+2\beta-\frac{\pi^{2}}{2}, \qquad 
\overline b_{2}(\beta) = b_{2}(\beta)+\frac{1}{2}\beta, \qquad 
\overline b_{3}(\beta) = b_{3}(\beta)+\frac{1}{8}+\frac{1}{48} \beta,\qquad \dots .
\ee
The explicit form of the first two functions $\overline{b}_{n}(\beta)$ is  
\ba
\la{5.11}
\overline b_{1}(\beta) &= 2\beta-\frac{\pi^{2}}{2}+\frac{\pi^{2}}{\beta}-\frac{2\pi^{2}}{\beta^{2}}+\frac{4\pi^{2}+\frac{\pi^{4}}{12}}{\beta^{3}}-\frac{8\pi^{2}+\frac{2\pi^{4}}{3}}{\beta^{4}}+\cdots, \\
\la{5.12}
\overline b_{2}(\beta) &= \frac{\beta }{2}+\frac{-3-\frac{\pi ^2}{4}}{\beta }+\frac{3+\frac{5 \pi ^2}{4}}{\beta^2}
+\frac{-4-\frac{29 \pi ^2}{6}-\frac{\pi^4}{16}}{\beta^3}+\frac{6+16 \pi ^2+\frac{19 \pi ^4}{24}}{\beta^4}+\cdots,
\ea
with similar expressions for the next ones.
The functions $\overline{b}_{n}(\beta)$ can be compared with the saddle-point analysis 
of Section~\ref{sec:saddle}, and one finds full agreement. For instance, it is easy to see that 
\be
\overline b_{1}(\beta) = -2u^{2}+4u\tan u = \sigma_{0}(u), 
\ee
where $u=u(\beta)$ from (\ref{3.3}). This can be checked by replacing the large-$\beta$ solution of (\ref{3.3}) \footnote{
Notice that $\overline b_{1}(\beta)$ is even in $\beta$, just as the partition function itself is. This is hidden in (\ref{5.11}), 
which is an expansion for $\beta\to +\infty$.}
\be
u = \frac{\pi}{2}-\frac{\pi}{\beta}+\frac{2\pi}{\beta^{2}}-\frac{\pi(24+\pi^{2})}{6 \beta^{3}}+\cdots.
\ee 
Similarly, the $O(\l^{0})$ correction in (\ref{3.4}), \ie the function $\sigma_{1}(u)$ in (\ref{3.5}) is 
\ba
\la{5.15}
\sigma_{1}(u) &= u\tan u +\log\cos u-\frac{1}{2}\log(1+u\tan u) \lp
 = \frac{\beta}{2}+\frac{1}{2}\log 2+\log \pi-\frac{3}{2}\log\beta
-\frac{3+\frac{\pi^{2}}{4}}{\beta}+\frac{3+\frac{5\pi^{2}}{4}}{\beta^{2}}+\cdots\lp
= \frac{1}{2}\log 2+\log \pi-\frac{3}{2}\log\beta+\overline b_{2}(\beta),
\ea
and this shows that (\ref{3.4}) is in agreement with (\ref{5.7}), taking into account the $O(\l^{0})$ contribution from 
the expansion of the first term
\be
\log\frac{\pi\sqrt 2(1-e^{-\l})^{3/4}}{\l^{3/4}}= \frac{1}{2}\log 2+\log \pi-\frac{3}{8}\l+O(\l^{2}).
\ee

\subsubsection{Schwarzian limit}
\la{sec:schwarzian}

One can consider the Schwarzian (triple-scaling) limit \cite{Okuyama:2025fhi,Lin:2022rbf}, defined by sending 
$\l\to0$ and $\beta\to\infty$ with the combination $\bjt\equiv2\l\beta$ held fixed. In this limit,
\be
\la{5.17}
\frac{2\beta}{\sqrt{\l(1-e^{-\l})}} = \frac{\bjt}{\l^{2}}+O(\l^{-1}),
\ee
and we get $\bjt=2\l\beta$. In this limit we get
\ba
\log Z &= \frac{1}{\l^{2}}\bjt+\bigg(-\frac{\pi^{2}}{2}+\frac{\bjt}{4}\bigg)\frac{1}{\l}+\frac{3}{2}\log\l\lp
-\frac{3}{2}\log\bjt
+\log(4\pi)+\frac{2\pi^{2}}{\bjt}+\frac{\bjt}{96}\lp
+\bigg[-\frac{1}{4}-\frac{6}{\bjt}\bigg(1+\frac{\pi^{2}}{12}\bigg)-\frac{8\pi^{2}}{\bjt^{2}}-\frac{\bjt}{384}\bigg]\l+O(\l^{2}).
\ea
The first line contains the JT partition function terms coming from ground state renormalization and measure -- this is the statement, made precise in \cite{Okuyama:2025fhi}, 
that the DSSYK energy scale $E_{0}=2/\sqrt{\l(1-e^{-\l})}$ itself has a $\l$-expansion $E_{0}=\frac2\l+\frac12+\frac{\l}{24}+\cdots$, and $\beta E_{0}$ 
generates precisely this tower of $\bjt$-proportional terms order by order. The second line contains the Schwarzian disk contribution $\frac{2\pi^{2}}{\bjt}-\frac{3}{2}\log \bjt$ -- 
matching the one-loop-exact Schwarzian partition function of \cite{Stanford:2017thb,Mertens:2017mtv}, and reproducing the leading semiclassical 
matching of \cite{Maldacena:2016upp,Goel:2023svz} -- plus a constant; terms linear in $\bjt$ at all orders in $\l$ arise from the expansion of the ground state energy, as above. 
The last line similarly contains the first quantum corrections. More terms at higher orders in $\l$ are similar and contain a finite number of $1/\bjt^{n}$ terms.

The Schwarzian limit discussed above has recently been re-derived starting directly from the bilocal-Liouville path integral of DSSYK, 
by identifying the soft modes as reparametrizations of suitably `twisted' time coordinates \cite{Berkooz:2024ifu}. 
The leading low-temperature free energy found there, their eq.~(2.14), reproduces the terms collected in (\ref{5.9})--(\ref{5.11}) once $\overline b_{1}(\beta)$ and $\overline b_{2}(\beta)$ 
are expanded at large $\beta$, providing a check of our expansion by an independent, non-modular method. 
The same reference identifies further bilocal-Liouville saddles, associated with reparametrizations winding several times around the thermal circle; 
we return to this comparison in the concluding Section~\ref{sec:conclusions}.

\subsection{Non perturbative corrections}

In this section, we extract the non-perturbative corrections to the partition function at low temperature by 
systematically exploiting the dualities (\ref{5.3}) and (\ref{5.4}).
This gives the following split representation of the partition function
\ba
\la{5.19}
\log Z(\beta) &= [\log Z(\beta)]_{\rm pert}+[\log Z(\beta)]_{\rm NP},
\ea
where the non-perturbative part (NP) is a power series in $\wt q$ with temperature dependent coefficients
\ba
\la{5.20}
[\log Z(\beta)]_{\rm NP} &= L_{1}(\beta)\,\wt q+L_{2}(\beta)\, \wt q^{2}+\cdots.
\ea

\subsubsection{Leading order}
Let us focus on the leading non-perturbative correction $O(\wt q)$. It is obtained by replacing
\bea
E_{2}(\wt q) &= 1-24\wt q+O(\wt q^{2}), \qquad
E_{4}(\wt q) = 1+240\wt q+O(\wt q^{2}), \\ 
E_{6}(\wt q) &= 1-504\wt q+O(\wt q^{2}), \qquad
(\wt q; \wt q)_{\infty}^{3} = 1-3\wt q+O(\wt q^{2}).
\eea
The leading term in (\ref{5.20}) may be written at low temperature as 
\ba
\la{5.22}
L_{1}(\beta) &= -3+\frac{d_{1}^{\rm NP}(\l)}{\beta}+\frac{d_{2}^{\rm NP}(\l)}{\beta^{2}}+\cdots,
\ea
where the exact expressions for the functions $d_{n}^{\rm NP}(\l)$
are
\bea
d_{1}^{\rm NP}(\l) &= -\frac{48\pi^{2}}{\l^{2}}\frac{1}{2}\sqrt{\l(1-e^{-\l})}, \\
d_{2}^{\rm NP}(\l) &= \bigg(-\frac{384\pi^{4}}{\l^{4}}+\frac{192\pi^{2}}{\l^{3}}-\frac{24\pi^{2}}{\l^{2}}\bigg)\frac{1}{4}\l(1-e^{-\l}),
\eea
and so on. Contrary to what happened in (\ref{5.7}), we cannot collect the various powers of $\l$
because $d_{n}^{\rm NP}(\l)\sim \l^{-n}$ for $\l\to 0$. In this case, the structure of the small-$\l$ expansion takes the natural form 
\ba
\la{5.24}
L_{1}(\beta)  &= f_{1,0}(y)+ f_{1,1}(y)\,\l+f_{1,2}(y)\,\l^{2}+\cdots,
\ea
where we introduced the fixed combination
\be
\la{5.25}
y = \frac{24\pi^{2}}{\l\beta}.
\ee
This implies that we are focusing on the large-$\beta$ and small-$\l$ regime with fixed $\beta\l$. From (\ref{4.22}), we may read the expansion of the functions $f_{n,p}(y)$ in (\ref{5.24})
and we find
\ba
\la{5.26}
f_{1,0}(y) &= -3-y-\frac{y^2}{6}-\frac{y^3}{54}-\frac{y^4}{648}-\frac{y^5}{9720}-\frac{y^6}{174960}+\cdots , \\
f_{1,1}(y) &=\frac{y}{4}+\frac{(1+\pi ^2) y^2}{12 \pi ^2}+\frac{(2+\pi ^2) y^3}{72 
\pi ^2}+\frac{(3+\pi ^2) y^4}{648 \pi ^2}+\frac{(4+\pi ^2) y^5}{7776 
\pi ^2}+\frac{(5+\pi ^2) y^6}{116640 \pi ^2}+\cdots, \\
f_{1,2}(y) &= -\frac{5 y}{96}-\frac{(15+8 \pi ^2) y^2}{288 \pi ^2}-\frac{(24+89 \pi 
^2+22 \pi ^4) y^3}{3456 \pi ^4}-\frac{(108+177 \pi ^2+28 \pi ^4) 
y^4}{31104 \pi ^4}\lp
-\frac{(144+147 \pi ^2+17 \pi ^4) y^5}{186624 \pi ^4}-\frac{(15+11 \pi ^2+\pi ^4) y^6}{139968 \pi ^4}+\cdots,
\ea
and so on.
One checks that a consistent resummation is given by
\bea
\la{5.29}
f_{1,0}(y) &= -3 e^{y/3}, \qquad f_{1,1}(y) = \frac{1}{4}e^{y/3} y \bigg(1+\frac{y}{3\pi^{2}}\bigg),\\
f_{1,2}(y) &= -\frac{1}{96} e^{y/3} y \bigg[5+y
+\frac{y (180+29 y)}{36 \pi ^2}
+\frac{y^2 (6+y)}{9 \pi ^4}
\bigg].
\eea
A rigorous calculation of the functions $f_{n,p}$ proving the closed formulas in (\ref{5.29}) is presented in Appendix~\ref{app:resum}.

\subsubsection{Next-to-leading order}

The next-to-leading order correction $O(\wt q^{2})$ is computed similarly starting from 
\bea
E_{2}(\wt q) &= 1-24\wt q-72\wt q^{2}+O(\wt q^{3}), \qquad
E_{4}(\wt q) = 1+240\wt q+2160\wt q^{2}+O(\wt q^{3}), \\ 
E_{6}(\wt q) &= 1-504\wt q-16632\wt q^{2}+O(\wt q^{3}), \qquad
(\wt q; \wt q)_{\infty}^{3} = 1-3\wt q+0\cdot \wt q^{2}+O(\wt q^{3}).
\eea
The correction $L_{2}(\beta)$ in (\ref{5.20}) has the same structure as in (\ref{5.24}, \ref{5.25}), \ie
\ba
L_{2}(\beta)  &= f_{2,0}(y)+ f_{2,1}(y)\,\l+
f_{2,2}(y)\,\l^{2}+\cdots.
\ea
From (\ref{4.22}), we can write the explicit small-$y$ series expansion of the functions $f_{2,n}(y)$; their exact resummation can again be determined
by the methods of Appendix~\ref{app:resum}.
The first three functions are 
\bea
f_{2,0}(y) &= -\frac{9}{2}e^{2y/3}, \qquad
f_{2,1}(y) = \frac{3}{4} e^{2y/3}y\bigg(1+\frac{y}{3\pi^{2}}\bigg), \\
f_{2,2}(y) &= -\frac{5}{32}e^{2y/3}y\bigg(1+\frac{2y}{5}+\frac{y(180+53y)}{180\pi^{2}}+\frac{2y^{2}(3+y)}{45\pi^{4}}\bigg).
\eea

\subsection{Higher-order non-perturbative correction and modular resummation}
\la{sec:higher-order}

The pattern continues at higher order. In particular, the first three functions $f_{n,0}$, $f_{n,1}$, $f_{n,2}$ for $n=3,4$, \ie the contributions
at $O(\wt q^{3})$ and $O(\wt q^{4})$, are respectively
\bea
f_{3,0}(y) &= -4e^{y}, \qquad
f_{3,1}(y) = e^{y}y\bigg(1+\frac{y}{3\pi^{2}}\bigg), \\
f_{3,2}(y) &= -\frac{1}{24}e^{y}y\bigg(5+3 y+\frac{y (180+67 y)}{36\pi^{2}} +\frac{y^2 (2+y)}{3\pi^{4}}  \bigg),
\eea
and
\bea
f_{4,0}(y) &= -\frac{21}{4}e^{4y/3}, \qquad
f_{4,1}(y) = \frac{7}{4} e^{4y/3}y\bigg(1+\frac{y}{3\pi^{2}}\bigg), \\
f_{4,2}(y) &= -\frac{7}{96}e^{4y/3}y\bigg(5+4 y+\frac{y (1260+587 y) }{252\pi^{2}} +\frac{2y^2 (3+2 y)}{9\pi^{4}}\bigg).
\eea
We computed the three functions $f_{n,0}$, $f_{n,1}$, $f_{n,2}$ for $n$ up to 20 and find that the following 
general expressions hold
\bea
\la{5.35}
f_{n,0}(y) &= -\frac{3\sigma_{1}(n)}{n}\,e^{ny/3}, \qquad
f_{n,1}(y) = \frac{1}{4}\sigma_{1}(n)\, e^{ny/3}y\bigg(1+\frac{y}{3\pi^{2}}\bigg), \\
f_{n,2}(y) &= -\frac{1}{96}\sigma_{1}(n)\,e^{ny/3}y\bigg(5+n y+\frac{y (180+34ny-5y \frac{\sigma_{3}(n)}{\sigma_{1}(n)}) }{36\pi^{2}} +\frac{y^2 (6+ny)}{9\pi^{4}}\bigg),
\eea
where $\sigma_{m}(n)$ is the sum of $m$-th powers of all divisors of $n$. We now define
\be
\la{5.36}
\Lambda = e^{y/3}\wt q = \exp\bigg[-\frac{4\pi^{2}}{\l}\bigg(1-\frac{2}{\beta}\bigg)\bigg],
\ee
and use the sums, \cf Appendix~\ref{app:sums},
\bea
\la{5.37}
\sum_{n=1}^{\infty}\frac{\sigma_{1}(n)}{n}\Lambda^{n} &= -\log(\Lambda; \Lambda)_{\infty}, \qquad\qquad\qquad
\sum_{n=1}^{\infty}\sigma_{1}(n)\Lambda^{n} = \frac{1}{24}(1-E_{2}(\Lambda)), \\
\sum_{n=1}^{\infty}\sigma_{1}(n)\,n\,\Lambda^{n} &= -\frac{1}{288}[E_{2}^{2}(\Lambda)-E_{4}(\Lambda)], \qquad
\sum_{n=1}^{\infty}\sigma_{3}(n)\,\Lambda^{n} = \frac{1}{240}(E_{4}(\Lambda)-1).
\eea
Substituting them in the general expressions (\ref{5.35})
gives the following full resummation of the non-perturbative contributions
\ba
\la{5.38}
[\log Z(\beta)]_{\rm NP} &= 3\,\log(\Lambda; \Lambda)_{\infty}+\frac{1-E_{2}}{96}y\,\bigg(1+\frac{y}{3\pi^{2}}\bigg)\, \l \lp
-\frac{y}{27648}\bigg[60(1-E_{2})-(E_{2}^{2}-E_{4})\,y
+\frac{y}{18\pi^{2}} \bigg(1080(1-E_{2})+(3-17E_{2}^{2}+14E_{4})\, y\bigg)\lp
 +\frac{y^2}{9\pi^{4}}\bigg(72(1-E_{2})-(E_{2}^{2}-E_{4})\, y\bigg)
\bigg]\,\l^{2}+\cdots
\ea
where
\be
E_{2n}\equiv E_{2n}(\Lambda).
\ee
Remarkably, this expression depends on the non-perturbative parameter $\Lambda$ via quasi-modular Eisenstein series.

The leading, $O(\l^{0})$ term in (\ref{5.38}), $3\log(\Lambda;\Lambda)_{\infty}$, has in fact a simple origin: it is just the $S$-duality 
transform (\ref{5.4}) of the leading large-$x$ term $S_{0}(q)=(q;q)_{\infty}^{3}$ of the Bessel expansion (\ref{4.22}), and coincides with 
the result of \cite{Okuyama:2025fhi}. The genuinely new content of (\ref{5.38}) is the $O(\l)$ and $O(\l^{2})$ corrections, which require 
the subleading terms in (\ref{4.22}) and could not have been obtained from the leading Bessel asymptotic alone.

\subsubsection{Sparsity of $Z(\beta)$ versus $\log Z(\beta)$}

Although $[\log Z(\beta)]_{\rm NP}$ in (\ref{5.38}) receives contributions at every power $\Lambda^{n}$, the partition function 
$Z(\beta)$ itself is far more constrained. The point is that $F(x)$ in (\ref{4.20}) is a \emph{linear} sum over $k$, 
$F(x)=\sqrt{2/\pi}\,x^{-3/2}e^{x}\sum_{k}\phi_{k}(q)\,x^{-k}$, with no products or exponentials mixing different $k$'s -- and each 
$\phi_{k}(q)$ is generated from $\phi_{0}=S_{0}(q)\equiv(q;q)_{\infty}^{3}=\sum_{j=0}^{\infty}(-1)^{j}(2j+1)q^{j(j+1)/2}$ by the ladder 
recursion (\ref{6.5}), an iterated application of $\sD=q\,d/dq$. Since $\sD$ acts diagonally on powers of $q$, it can never introduce 
powers of $q$ different from those in $S_{0}$ itself: every $\phi_{k}(q)$, for any $k$, contains exactly the same $q^{j(j+1)/2}$ powers with 
exponents equal to triangular integers.
As a linear combination of such terms, $F(x)$ -- and hence $Z(\beta)$ -- inherits this same property, order by 
order in $1/x$, before any expansion in $\l$ is even performed.

This survives $S$-duality intact: the transformation rules (\ref{5.3})--(\ref{5.4}) map each $E_{2m}(q)$ and $S_{0}(q)$ into a 
$q$-independent ($\l$-dependent) prefactor times the same object evaluated at $\wt q$, so each $\phi_{k}(\wt q)$ remains supported on 
the triangular numbers after duality. Consequently, the entire non-perturbative sector of $Z(\beta)$ -- to all orders in $\l$, not 
just the orders where $[\log Z(\beta)]_{\rm NP}$ happens to be tractable -- has non-vanishing $\wt q^{n}$ contributions only at 
$n=j(j+1)/2$, $j=0,1,2,\ldots$: exactly the winding/conical-defect saddle spectrum of \cite{Berkooz:2024ifu} discussed in 
Section~\ref{sec:conclusions}. The appearance of all powers of $\wt q$ in $[\log Z(\beta)]_{\rm NP}$ is simply an artifact of the logarithm: it is 
$Z(\beta)$, not its logarithm, that is the natural object to compare against the saddle spectrum. As an illustration, the first 
two instances of this mechanism are
\be
S_{0}E_{2} = \sum_{j=0}^{\infty}(-1)^{j}(2j+1)^{3}\Lambda^{j(j+1)/2}, \qquad
S_{0}(5E_{2}^{2}-2E_{4}) = 3\sum_{j=0}^{\infty}(-1)^{j}(2j+1)^{5}\Lambda^{j(j+1)/2},
\ee
which are exactly the combinations appearing in the $O(\l)$ and $O(\l^{2})$ terms of $Z_{\rm NP}(\beta)$ 
once (\ref{5.38}) is exponentiated.

\section{A differential equation for the partition function}
\la{sec:modular}

While the previous sections presented a direct analysis of the low-temperature expansions completed with explicit non-perturbative corrections, 
in this section we present a further structural property of the partition function.

In fact, the quasi-modular expansion (\ref{4.22}) can be shown to be the large-$x$ solution of a single, closed differential equation relating $\sD$ and $x$-derivatives of $F(x)$.
To derive it, we begin by writing $H(x)\equiv \sum_{n}(-1)^{n}(2n+1)q^{n(n+1)/2}I_{2n+1}(x) = \tfrac{x}{2}F(x)$, cf.\ (\ref{2.7}). Acting with $\sD=q\,d/dq$ pulls down $\tfrac{n(n+1)}{2}$ on each term,
\be
\sD H = \frac18\sum_{n}(-1)^{n}(2n+1)q^{n(n+1)/2}\cdot 4n(n+1)\,I_{2n+1}(x).
\ee
Now use $4n(n+1)=\nu^{2}-1$ for $\nu=2n+1$, together with Bessel's equation $\nu^{2}I_{\nu}=x^{2}I_{\nu}''+xI_{\nu}'-x^{2}I_{\nu}$. Since this differential operator in $x$ 
is the same for every $n$, we get
\be
\sD H = \frac18\Big[x^{2}\partial_{x}^{2}+x\partial_{x}-x^{2}-1\Big]H.
\ee
Converting back to $F=2H/x$ gives the closed equation
\be
\la{6.3}
8\,\sD F(x) = \Big[x^{2}\partial_{x}^{2}+3x\partial_{x}-x^{2}\Big]F(x),
\ee
which is easily checked against the explicit $q$-series of $F(x)$ order by order in $x$ and $q$ presented in (\ref{4.22}).
However, equation (\ref{6.3}) is more general, and exact. 

It is straightforward to show that the large-$x$ expansion of (\ref{6.3}) generates (\ref{4.22}).
Substituting the large-$x$ ansatz $F=\sqrt{2/\pi}\,x^{-3/2}e^{x}\Psi(x)$, $\Psi(x)=\sum_{k\geq0}\phi_{k}\,x^{-k}$, into (\ref{6.3}) reduces it to
\be
8\,\sD\Psi = x^{2}\Psi''+2x^{2}\Psi'-\frac34\,\Psi,
\ee
and matching powers of $x^{-k}$ gives the two-term recursion
\be
\la{6.5}
\phi_{k+1} = -\frac{1}{8(k+1)}\Big[32\sD+3-4(k+1)k\Big]\phi_{k}, \qquad \phi_{0}=1.
\ee
This is exactly the ladder recursion $T_{k}=[32\sD+3-4k(k-1)]T_{k-1}$, \cf (\ref{4.20}), with $\phi_{k}=\frac{(-1)^{k}}{k!\,8^{k}}T_{k}$. 
In other words, the quasi-modular ring closure of Section~\ref{sec:modular-structure} 
is not an independent fact about $E_{2},E_{4},E_{6}$: it is simply what (\ref{6.3}) becomes 
when solved order-by-order in $1/x$, using the fact that $\sD$ preserves the ring $\mathbb C[E_{2},E_{4},E_{6}]$.

Structurally, (\ref{6.3}) is a modular-heat-kernel-type equation -- first order in $\tau$ (via $\sD$), second order in $x$ -- entirely analogous to the Jacobi theta
function's heat equation $\partial_{\tau}\vartheta = \frac{1}{4\pi i}\partial_{z}^{2}\vartheta$.

\paragraph{Equation for $Z(\beta)$} Equation (\ref{6.3}) treats $x$ and $\l$ (through $q=e^{-\l}$) as independent variables, with $\sD=-\partial_{\l}|_{x}$. 
Since the physical partition function depends on $\l$ \emph{and} $\beta$ through $x=x(\beta,\l)$ in (\ref{2.6}), it is useful to rewrite (\ref{6.3}) directly in terms of $(\beta,\l)$, i.e.\ for $Z(\beta,\l)\equiv F(x(\beta,\l),\l)$.
As usual, for simplicity we will omit the argument $\l$, which is always understood. A short calculation gives 
\be
\la{6.6}
-8\,\partial_{\l}Z = \beta^{2}\partial_{\beta}^{2}Z
+\bigg[3+\frac4\l+\frac{4}{e^{\l}-1}\bigg]\beta\,\partial_{\beta}Z
-\frac{4 \beta^{2}}{\l(1-e^{-\l})}\,Z.
\ee
We now show how the differential equation (\ref{6.6}) reproduces the perturbative and non-perturbative parts of the 
low-temperature expansion of the partition function in the semiclassical limit.

\subsection{Determining the functions $\overline{b}_{n}(\beta)$ from the differential equation}

The equation (\ref{6.6}) written for $\log Z$ is 
\be
\la{6.7}
-8\,\partial_{\l}\log Z = \beta^{2}\partial_{\beta}^{2}\log Z+\beta^{2}(\partial_{\beta}\log Z)^{2}
+\bigg[3+\frac4\l+\frac{4}{e^{\l}-1}\bigg]\beta\,\partial_{\beta}\log Z
-\frac{4 \beta^{2}}{\l(1-e^{-\l})}.
\ee
We may substitute (\ref{5.9}), \ie the semiclassical expansion of the perturbative contribution, and, expanding in small $\l$, we get the equations
\ba
& -8 \overline{b}_1(\beta )+\beta  (-4 \beta +\overline{b}_1'(\beta ) 
(8+\beta  \overline{b}_1'(\beta ))) = 0, \notag \\
\la{6.8}
& -2 (6+\beta ^2)+\beta  (8 \overline{b}_2'(\beta )+2 \overline{b}_1'(\beta ) (-1+\beta 
 \overline{b}_2'(\beta ))+\beta  \overline{b}_1''(\beta ))=0, \\
 & -\frac{3}{4}-\frac{\beta ^2}{3}+8 \overline{b}_3(\beta )+\frac{1}{3} 
\beta  \overline{b}_1'(\beta ) (1+6 \beta  \overline{b}_3'(\beta ))+\beta [
 \overline{b}_2'(\beta ) (-2+\beta  \overline{b}_2'(\beta ))+8 
\overline{b}_3'(\beta )+\beta  \overline{b}_2''(\beta )] =0, \notag
\ea
and so on. These equations encode the exact solution to the saddle-point method, \cf (\ref{3.5}). It is 
instructive to begin with the determination of $\overline{b}_{1}(\beta)$.
Let us introduce
the lighter notation $f(\beta)\equiv \overline{b}_{1}(\beta)$ and solve
\be
\la{6.9}
\beta^{2}f'^{2}+8\beta f'-8f-4\beta^{2}=0.
\ee
We write it in Lagrange form 
\be
\la{6.10}
f = \beta f'+\frac{\beta^{2}}{8}(f'^{2}-4).
\ee
The so-called singular solutions with constant $f'$ are readily found to be $f(\beta) = \pm 2 \beta$. These are ruled out by our explicit expansions showing that $f$ is not simply linear in $\beta$.
The other solutions are obtained by introducing $p=f'$ and considering $\beta=\beta(p)$. Differentiating (\ref{6.10}) with respect to $\beta$ gives
\be
p = \frac{d}{d\beta}\bigg[\beta p+\frac{\beta^{2}}{8}(p^{2}-4)\bigg]\qquad \to \qquad
p = p+\beta\frac{dp}{d\beta}+\frac{\beta}{4}(p^{2}-4)+\frac{\beta^{2}}{4}p\,\frac{dp}{d\beta},
\ee
and thus we get the simple equation 
\be
\frac{d\beta}{dp}=\frac{4+p \beta}{4-p^{2}}.
\ee
Let us change variables and write $p=2\sin u$. We get 
\be
\frac{d\beta}{du}=\frac{2+\beta\sin u}{\cos u}\qquad\to \qquad  \beta(u) = \frac{2u+c}{\cos u},
\ee
where $c$ is an integration constant, which we set to zero by symmetry $\beta\to -\beta$. Substituting in (\ref{6.10}) gives
\be
f = \frac{2u}{\cos u}2\sin u+\bigg(\frac{2u}{\cos u}\bigg)^{2}(4\sin^{2}u-4) = -2u^{2}+4u\tan u,
\ee
in agreement with (\ref{3.5}). The determination of the higher functions $\overline{b}_{n>1}$ is much easier since their equations are linear, \cf (\ref{6.8}).
For instance, one can check that $\sigma_{1}$ is reproduced using (\ref{5.15}).

\subsection{Non-perturbative corrections from the differential equation}

As an example, let us derive from (\ref{6.7}) the leading-order and next-to-leading order non-perturbative corrections $f_{1,0}(y)$ and $f_{1,1}(y)$.
Let us substitute in (\ref{6.7}) the expansion 
\ba
\log Z &= \log\frac{\pi\sqrt 2\, (1-e^{-\l})^{3/4}}{\l^{3/4}}-\frac{3}{2}\log\beta 
+\frac{\overline b_{1}(\beta)}{\l}+\overline b_{2}(\beta)+\overline b_{3}(\beta)\,\l+\cdots\lp
+(f_{1,0}(y)+f_{1,1}(y)\l+\cdots)\, \wt q+O(\wt q^{2}),
\ea
with $y$ as in (\ref{5.25}).
The leading term in small $\lambda$ is 
\be
\la{6.16}
\frac{32\pi^{2}}{\l^{2}}\bigg[f_{1,0}(y)-\frac{3}{2}\overline b_{1}'\, \bigg(\frac{24\pi^{2}}{y\l}\bigg)\,f_{1,0}'(y)\bigg]+O(\l^{-1}).
\ee
At small $\l$, we need the value of $\overline{b}_{1}'$ at large argument. From  (\ref{5.11}), we get $\overline{b}_{1}'(\beta)\to 2$ for $\beta\to +\infty$
and thus the vanishing of (\ref{6.16}) is equivalent to a simple differential equation for $f_{1,0}$ that gives
\be
\la{6.17}
f_{1,0}(y) = C e^{y/3},
\ee
where the integration constant is $C=-3$, fixed by the first term in (\ref{5.26}). In this way, we show that the master equation (\ref{6.7}) reproduces the leading function $f_{1,0}$.
The vanishing of the order $O(1/\l)$ contributions to the differential equation is similarly computed to give
\be
\la{6.18}
f_{1,1}(y)-\frac{1}{2\pi^{2}}\,y\,f_{1,0}'(y)-\frac{3}{2}\bigg[f_{1,1}'(y)\overline b_{1}'\, \bigg(\frac{24\pi^{2}}{y\l}\bigg)+f_{1,0}'(y)\overline b_{2}'\, \bigg(\frac{24\pi^{2}}{y\l}\bigg)\bigg] = 0.
\ee
We now use again the large-$\beta$ limit $\overline{b}_{1}'(\beta)\to 2$ and, from (\ref{5.12}), the limit $\overline{b}_{2}'(\beta)\to 1/2$. Substituting into (\ref{6.18})
the solution we found for $f_{1,0}(y) = -3e^{y/3}$, we get the previous expression for $f_{1,1}(y)$ in (\ref{5.29}),
with the integration constant fixed by matching to the small-$y$ series in (\ref{5.26}).

In the same way, one may derive the other functions and higher non-perturbative terms by further expanding in $\wt q$.

\subsection{High-temperature expansion from the differential equation}
\la{sec:high-T-from-ODE}

The differential equation (\ref{6.3}) is equally informative in the opposite, high-temperature regime, where $F(x)$ is 
analytic in $x$ and $Z(\beta)$ is given by the convergent small-$x$ expansion of Section~\ref{sec:saddle}. Substituting the ansatz
\be
\la{6.19}
F(x) = \sum_{k=0}^{\infty} c_{k}(q)\, x^{2k}, \qquad c_{0}=1,
\ee
into (\ref{6.3}) and matching powers of $x^{2k}$ gives the linear recursion
\be
\la{6.20}
4k(k+1)\, c_{k}(q) - 8\,\sD c_{k}(q) = c_{k-1}(q).
\ee
Unlike the large-$x$ ladder (\ref{6.5}), this recursion is first order in $\sD$ and only fixes $c_{k}(q)$ up to an integration 
constant at each step. The missing input is supplied by the elementary Bessel identity
\be
\la{6.21}
F(x)\big|_{q=1} = \frac{2}{x}\sum_{n=0}^{\infty}(-1)^{n}(2n+1)\,I_{2n+1}(x) = 1,
\ee
valid for every $x$, which fixes $c_{k}(1)=0$ for all $k\geq1$. Together, (\ref{6.20}) and (\ref{6.21}) determine every 
$c_{k}(q)$ uniquely and reproduce the coefficients of Section~\ref{sec:saddle} (e.g.\ $c_{1}=(1-q)/8$, $c_{2}=(2-3q+q^{3})/384$, 
and so on) entirely from the differential equation, with no further input.

This construction also proves a structural fact about these coefficients that Section~\ref{sec:saddle} otherwise only checks order 
by order: $c_{k}(q)$ is divisible by $(1-q)^{k}$. Writing $q=1+\epsilon$ and $c_{k}(\epsilon)=\sum_{j}a_{j}\epsilon^{j}$, the 
recursion (\ref{6.20}) reads, order by order in $\epsilon$,
\be
(j+1)\,a_{j+1} = \bigg(\frac{k(k+1)}{2}-j\bigg)a_{j} - \frac18\, b_{j},
\ee
where $b_{j}$ is the $\epsilon^{j}$ coefficient of $c_{k-1}(\epsilon)$. If $c_{k-1}$ already vanishes to order $k-1$ (i.e.\ 
$b_{j}=0$ for $j<k-1$, the inductive hypothesis), this becomes a pure two-term recursion for $j<k-1$, and the single condition 
$a_{0}=c_{k}(1)=0$ then cascades through $a_{1},\ldots,a_{k-1}$, forcing them all to vanish. Starting from $c_{1}\propto(1-q)^{1}$, 
induction on $k$ then shows $c_{k}(q)\propto(1-q)^{k}$ for every $k$.

The same strategy determines $\log F(x)=\sum_{m\geq1}g_{m}(q)\,x^{2m}$, which obeys the nonlinear analogue of (\ref{6.20}),
\be
\la{6.23}
4m(m+1)\,g_{m}(q) - 8\,\sD g_{m}(q) = \delta_{m,1} - \sum_{k=1}^{m-1}4k(m-k)\,g_{k}(q)\,g_{m-k}(q),
\ee
again with $g_{m}(1)=0$ from (\ref{6.21}). This reproduces the coefficients quoted in Section~\ref{sec:saddle} (e.g.\ 
$g_{2}=-(1-q)^{3}/384$, $g_{3}=(1-q)^{5}(5+q)/46080$), and the same cascade argument -- now using that each product 
$g_{k}g_{m-k}$ vanishes to order at least $(2k-1)+(2(m-k)-1)=2m-2$ by the inductive hypothesis -- shows that $g_{m}(q)$ 
is divisible by $(1-q)^{2m-1}$. This gives a first-principles derivation, directly from (\ref{6.3}), of the persistent 
$(1-q)^{n-1}x^{n}$ structure noted in Section~\ref{sec:saddle} and originally observed in \cite{Okuyama:2023iwu,Verg2013}.

\section{Conclusions}
\la{sec:conclusions}

Starting from the exact Bessel-function representation of $Z(\beta)$, we found that its expansion at large argument can 
be organized entirely in terms of the classical ring of quasi-modular Eisenstein series $E_{2},E_{4},E_{6}$, eq.~(\ref{4.22}). 
Exploiting the modular ($S$-duality) properties of this ring, we obtained the semiclassical expansion of $\log Z(\beta)$ at small $\l$, 
splitting it into a perturbative tower, systematic to all orders in both $1/\beta$ and $\l$, and a non-perturbative sector controlled by 
$\wt q=e^{-4\pi^{2}/\l}$. 

At each order $\wt q^{n}$, keeping systematically the subleading large-$\beta$ corrections neglected by the leading Bessel approximation of \cite{Okuyama:2025fhi}, 
we found that the non-perturbative correction resums, up to second order in $\l$, into the closed forms $f_{n,0}(y)$, $f_{n,1}(y)$ and $f_{n,2}(y)$, eq.~(\ref{5.35}), 
with coefficients built from the divisor functions $\sigma_{1}(n)$ and $\sigma_{3}(n)$ -- extending the $\beta$-independent, 
leading-order result of \cite{Okuyama:2025fhi}, $C_{n}=-3\sigma_{1}(n)/n$, to a genuine function of $y=24\pi^{2}/(\l\beta)$. 
Using the generating functions of $\sigma_{1}(n)$ and $\sigma_{3}(n)$ in terms of the quasi-modular Eisenstein series $E_{2},E_{4}$, 
the entire non-perturbative series up to this order resums into the closed expression (\ref{5.38}). 
We then showed that this structure -- perturbative and non-perturbative alike -- follows from a single, exact differential equation, eq.~(\ref{6.3}), 
relating a modular derivative and $x$-derivatives of the Bessel-function representation of $Z(\beta)$; both towers are recovered from it as solutions in the appropriate limits, 
with the divisor-sum coefficients entering as boundary data fixed by matching onto the quasi-modular expansion, rather than being derived from the differential equation itself.

Beyond providing a systematic, all-order completion of the semiclassical expansion, these results illustrate a broader point: 
the low-temperature behaviour of the DSSYK partition function, non-perturbative sector included, is 
controlled by essentially number-theoretic data -- Eisenstein series, divisor sums, and their modular transformation 
properties -- reflecting general modular-form technology rather than features specific to the detailed structure of the DSSYK Hamiltonian. 
The next natural question to ask is whether this structure admits a direct bulk interpretation.

To this end, it is tempting to compare the powers of $\Lambda$ defined in (\ref{5.36}) with the additional saddles of the bilocal-Liouville theory found in \cite{Berkooz:2024ifu}
in the triple-scaling limit discussed in Section~\ref{sec:schwarzian}. Besides the primary Schwarzian saddle discussed there -- 
whose free energy underlies their eq.~(2.14), matched against (\ref{5.9})--(\ref{5.11}) in Section~\ref{sec:schwarzian} -- one finds in \cite{Berkooz:2024ifu} a tower of 
further saddles labelled by $k=1,2,\dots$, corresponding to reparametrizations winding $2k+1$ times around the thermal circle (equivalently, a Schwarzian with 
conical defect $\alpha=2\pi(2k+1)$), with on-shell action given in their eq.~(3.33a) -- the same saddles have independently been 
discussed in the sine-dilaton gravity dual to DSSYK \cite{Blommaert:2024whf}. Denoting by $I_{k}$ the on-shell action of the 
$k$-th such saddle (with $J=1$) and by $I_{0}$ that of the primary saddle just mentioned, one finds
\be
\la{7.1}
I_{k}-I_{0} = \frac{2\pi^{2}k(k+1)}{\l}\bigg(1-\frac{2}{\beta}\bigg).
\ee
Comparing with the exponent $\frac{4\pi^{2}n}{\l}(1-2/\beta)$ from $\Lambda^{n}$, \cf (\ref{5.36}), the two expressions agree 
precisely when $n=k(k+1)/2$, \ie at the triangular numbers $n=1,3,6,10,\dots$ for $k=1,2,3,4,\dots$ -- and, as shown in 
Section~\ref{sec:higher-order}, these are exactly the exponents on which the non-perturbative sector of $Z(\beta)$ itself 
(as opposed to $[\log Z(\beta)]_{\rm NP}$) is supported, to all orders in $\l$. This exhausts the saddle spectrum found from 
both the classical bilocal-Liouville action and the exact DSSYK spectral density in \cite{Berkooz:2024ifu}, and the match with 
the non-perturbative sector of $Z(\beta)$ is exact, with no remaining exponents left over to be accounted for by additional 
contributions.
We stress that this exact match concerns only the spectrum of exponents; a comparison of the associated one-loop weights 
with $C_{n}=f_{n,0}(0)$ -- and, at subleading order in $\l$, with the functions $f_{n,1}(y)$ and $f_{n,2}(y)$ -- together 
with a first-principles derivation of the full divisor-sum structure from the saddles of \cite{Berkooz:2024ifu}, would be 
worth pursuing but lie beyond the scope of the present paper.

It would also be interesting to extend the present analysis to other observables, such as two-point functions: in the 
sine-dilaton gravity dual of DSSYK, the wormhole length -- and hence Krylov (spread) complexity -- can be extracted 
from the two-point function of a probe operator \cite{Heller:2024ldz}. The leading non-perturbative corrections to 
this quantity have already been identified in \cite{Alfinito:2026cky}, and it is natural to ask whether the 
quasi-modular structure and $S$-duality exploited here could play a similar organizing role.

\paragraph{Acknowledgements}

We thank B. Bajc and Y. Fu for useful comments.
MB is supported by the INFN grant GAST. EA is supported by the MUR project GINEVRA, prot.
2022BZYBWM.

\appendix
\section{Proof of the leading non-perturbative resummations}
\la{app:resum}

Here we derive, rather than guess-and-check, the closed forms in (\ref{5.29})
that appear in the small-$\l$, fixed-$\beta\l$ expansion of $[\log Z(\beta)]_{\rm NP}$.
We start by writing
\be
T_{k}(\l) \equiv \prod_{j=1}^{k}\big[32\sD + 3-4j(j-1)\big]\, S_{0} \ = \ A_{k}(\l) + B_{k}(\l)\,\wt q + O(\wt q^{2}),
\ee
so that $F(x)$ in (\ref{2.6})--(\ref{2.7}) is built from $T_k$ with $q=e^{-\l}$. Since $\wt q(\l) = e^{-4\pi^{2}/\l}$ is itself a function of $\l$, once everything is expressed in terms of $\l$,
the operator becomes an honest derivative,
\be
\sD = -\frac{d}{d\l},
\ee
and differentiating $T_{k-1}(\l)$ (smooth part \emph{and} $\wt q(\l)$ part together) gives two \emph{exact}, coupled recursions
\ba
\la{A.3}
A_{k} &= -32\, A_{k-1}' + c_{k}\, A_{k-1}, \\
\la{A.4}
B_{k} &= -32\Big[B_{k-1}' + \frac{4\pi^{2}}{\l^{2}}B_{k-1}\Big] + c_{k}\, B_{k-1},
\ea
with $c_{k} = 3-4k(k-1)$ and initial data fixed by $S_0 = A_0(\l)(1-3\wt q+O(\wt q^2))$, i.e.
\be
\la{A.5}
A_{0} = e^{\l/8-\pi^{2}/2\l}\Big(\frac{2\pi}{\l}\Big)^{3/2}, \qquad B_{0} = -3\,A_{0}.
\ee
We now introduce the two generating series in $x^{-1}$,
\ba
\la{A.6}
\Sigma^{(0)}(x) &= \sum_{k\geq0}\Phi_{k}^{(0)}x^{-k}, \qquad \Phi_{k}^{(0)} \equiv \frac{(-1)^{k}}{k!\,8^{k}}\frac{A_{k}}{A_{0}}, \\
\Sigma^{(1)}(x) &= \sum_{k\geq0}\Phi_{k}^{(1)}x^{-k}, \qquad \Phi_{k}^{(1)} \equiv \frac{(-1)^{k}}{k!\,8^{k}}\frac{B_{k}}{A_{0}}.
\ea
Then $F(x) = \sqrt{2/\pi}\,x^{-3/2}e^{x}A_{0}\big[\Sigma^{(0)}(x)+\wt q\,\Sigma^{(1)}(x)+O(\wt q^{2})\big]$, so that
\be
Z = Z_{\rm pert}\Big[1+\wt q\,\frac{\Sigma^{(1)}(x)}{\Sigma^{(0)}(x)}\Big]+O(\wt q^{2})
\quad\rightarrow\quad
[\log Z]_{\rm NP} = \frac{\Sigma^{(1)}(x)}{\Sigma^{(0)}(x)}\,\wt q.
\ee

\paragraph{Solving the recursions}

If we substitute in (\ref{A.3}) the ansatz
\be
A_{k} = A_{0}\bigg(-\frac{16\pi^{2}}{\l^{2}}\bigg)^{k}\bigg(1+A_{k}^{(1)}\l+A_{k}^{(2)}\l^{2}+\cdots\bigg),
\ee
we get simple recursions for the coefficients $A_{k}^{(1)}, A_{k}^{(2)}, \cdots$ and in particular
\ba
A_{k}^{(1)} &= A_{k-1}^{(1)}+\frac{1-4k}{\pi^{2}}, \\
\la{A.11}
A_{k}^{(2)} &= A_{k-1}^{(2)}-\frac{1}{8}A_{k}^{(1)}+\bigg(\frac{1}{8}+\frac{3-4k}{\pi^{2}}\bigg)\,A_{k-1}^{(1)}+\frac{3-12k+4k^{2}}{16\pi^{2}}.
\ea
The boundary condition is $A_{0}^{(n)}=0$ and thus
\be
A_{k}^{(1)} = -\frac{1}{\pi^{2}}k(1+2k), \qquad A_{k}^{(2)} = \frac{1}{48\pi^{4}}k(4k^{2}-1)(\pi^{2}+24(k-1)).
\ee
Replacing into (\ref{A.6}), writing $x$ in terms of $\beta$ using (\ref{2.6}), and finally replacing $\beta$ with $y$ as in (\ref{5.25}), we get 
\ba
\Sigma^{(0)} &= e^{y/24}\bigg[1+\bigg(-\frac{y}{96}-\frac{y(36+y)}{288\pi^{2}}\bigg)\,\l\lp
+\bigg(
\frac{y (40+y)}{18432}+\frac{y (5616+576 y+7 y^2)}{165888\pi^{2}}+\frac{y^2 (2160+120 y+y^2)}{165888\pi^{4}}
\bigg)\,\l^{2}+\cdots\bigg].
\ea
Similarly, if we substitute in (\ref{A.4})
\be
B_{k} = A_{0}\bigg(-\frac{144\pi^{2}}{\l^{2}}\bigg)^{k}\bigg(1+B_{k}^{(1)}\l+B_{k}^{(2)}\l^{2}+\cdots\bigg),
\ee
we find
\be
B_{k}^{(1)} = -\frac{1}{9\pi^{2}}k(1+2k), \qquad B_{k}^{(2)} = \frac{1}{1296\pi^{4}}k(4k^{2}-1)(3\pi^{2}+8(k-1)).
\ee
This gives
\ba
\Sigma^{(1)} &= e^{3y/8}\bigg[-3+\bigg(\frac{9y}{32}+\frac{3y(4+y)}{32\pi^{2}}\bigg)\,\l\lp
+\bigg(
-\frac{3 y (40+9 y)}{2048}-\frac{y (208+192 y+21 y^2)}{2048\pi^{2}}-\frac{y^2 (80+40 y+3 y^2)}{2048\pi^{4}}
\bigg)\,\l^{2}+\cdots\bigg].
\ea
Computing the ratio $\Sigma^{(1)}/\Sigma^{(0)}$ reproduces
\be
\frac{\Sigma^{(1)}}{\Sigma^{(0)}} = f_{1,0}(y)+f_{1,1}(y)\,\l+f_{1,2}(y)\,\l^{2}+\cdots,
\ee
with the functions in (\ref{5.29}). The procedure can be easily extended to higher orders. For instance, at order $O(\l^{3})$, 
one finds
\be
f_{1,3}(y) = e^{y/3}y\bigg[
\frac{27+15 y+y^2}{3456}+\frac{y (792+285 y+17 y^2)}{41472\pi^{2}}
+\frac{y^2 (1044+312 y+17 y^2)}{124416\pi^{4}}+\frac{y^3 (54+18 y+y^2)}{93312\pi^{6}}
\bigg].
\ee

\section{Sums involving the divisor functions}
\la{app:sums}
Let us discuss the proof of summation formulas (\ref{5.37}). The definition of Eisenstein sums in (\ref{4.2}) may also be written
\be
E_{2m} = 1-\frac{4m}{B_{2m}}\sum_{n=1}^{\infty}\sigma_{2m-1}(n)q^{n}.
\ee
This gives immediately
\be
\sum_{n=1}^{\infty}\sigma_{2m-1}(n)n^{p}q^{n} = \frac{B_{2m}}{4m}\mathscr D^{p}(1-E_{2m}),
\ee
that gives the sums in (\ref{5.37}) involving $E_{2m}$ upon using the Ramanujan differential identities (\ref{4.5}). To derive the first sum in (\ref{5.37}),
we use the identity -- following from the definition of $\sigma_{m}(n)$ -- 
\be
\sum_{n=1}^{\infty}\sigma_{m}(n)q^{n} = \sum_{n=1}^{\infty}\sum_{j=1}^{\infty}n^{m}q^{jn}.
\ee
In particular, for $m=1$, dividing both sides by $q$ and integrating in $q$, we get 
\be
\sum_{n=1}^{\infty}\frac{1}{n}\sigma_{1}(n)q^{n} = \sum_{n=1}^{\infty}\sum_{j=1}^{\infty}\frac{1}{j}q^{jn} = -\sum_{n=1}^{\infty}\log(1-q^{n}) = -\log(q;q)_{\infty}.
\ee

\bibliography{Krylov-Biblio}

\providecommand{\href}[2]{#2}\begingroup\raggedright\begin{thebibliography}{10}

\bibitem{Sachdev:1992fk}
S.~Sachdev and J.~Ye, \emph{{Gapless Spin Fluid Ground State in a Random,
  Quantum Heisenberg Magnet}},
  \href{https://doi.org/10.1103/PhysRevLett.70.3339}{\emph{Phys. Rev. Lett.}
  {\bfseries 70} (1993) 3339}
  [\href{https://arxiv.org/abs/cond-mat/9212030}{{\ttfamily
  cond-mat/9212030}}].

\bibitem{Sachdev:2010um}
S.~Sachdev, \emph{{Holographic Metals and the Fractionalized Fermi Liquid}},
  \href{https://doi.org/10.1103/PhysRevLett.105.151602}{\emph{Phys. Rev. Lett.}
  {\bfseries 105} (2010) 151602}
  [\href{https://arxiv.org/abs/1006.3794}{{\ttfamily 1006.3794}}].

\bibitem{Maldacena:2016hyu}
J.~Maldacena and D.~Stanford, \emph{{Remarks on the Sachdev-Ye-Kitaev Model}},
  \href{https://doi.org/10.1103/PhysRevD.94.106002}{\emph{Phys. Rev. D}
  {\bfseries 94} (2016) 106002}
  [\href{https://arxiv.org/abs/1604.07818}{{\ttfamily 1604.07818}}].

\bibitem{Maldacena:2016upp}
J.~Maldacena, D.~Stanford and Z.~Yang, \emph{{Conformal Symmetry and Its
  Breaking in Two Dimensional Nearly Anti-De-Sitter Space}},
  \href{https://doi.org/10.1093/ptep/pt$W_1$24}{\emph{PTEP} {\bfseries 2016}
  (2016) 12C104} [\href{https://arxiv.org/abs/1606.01857}{{\ttfamily
  1606.01857}}].

\bibitem{Jensen:2016pah}
K.~Jensen, \emph{{Chaos in AdS$_2$ Holography}},
  \href{https://doi.org/10.1103/PhysRevLett.117.111601}{\emph{Phys. Rev. Lett.}
  {\bfseries 117} (2016) 111601}
  [\href{https://arxiv.org/abs/1605.06098}{{\ttfamily 1605.06098}}].

\bibitem{Sarosi:2017ykf}
G.~S{\'a}rosi, \emph{{AdS$_{2}$ holography and the SYK model}},
  \href{https://doi.org/10.22323/1.323.0001}{\emph{PoS} {\bfseries Modave2017}
  (2018) 001} [\href{https://arxiv.org/abs/1711.08482}{{\ttfamily
  1711.08482}}].

\bibitem{Berkooz:2018jqr}
M.~Berkooz, M.~Isachenkov, V.~Narovlansky and G.~Torrents, \emph{{Towards a
  Full Solution of the Large $N$ Double-Scaled Syk Model}},
  \href{https://doi.org/10.1007/JHEP03(2019)079}{\emph{JHEP} {\bfseries 03}
  (2019) 079} [\href{https://arxiv.org/abs/1811.02584}{{\ttfamily
  1811.02584}}].

\bibitem{Mertens:2017mtv}
T.~G. Mertens, G.~J. Turiaci and H.~L. Verlinde, \emph{{Solving the Schwarzian
  via the Conformal Bootstrap}},
  \href{https://doi.org/10.1007/JHEP08(2017)136}{\emph{JHEP} {\bfseries 08}
  (2017) 136} [\href{https://arxiv.org/abs/1705.08408}{{\ttfamily
  1705.08408}}].

\bibitem{Engelsoy:2016xyb}
J.~Engels{\"o}y, T.~G. Mertens and H.~Verlinde, \emph{{An investigation of
  AdS$_{2}$ backreaction and holography}},
  \href{https://doi.org/10.1007/JHEP07(2016)139}{\emph{JHEP} {\bfseries 07}
  (2016) 139} [\href{https://arxiv.org/abs/1606.03438}{{\ttfamily
  1606.03438}}].

\bibitem{Mertens:2022irh}
T.~G. Mertens and G.~J. Turiaci, \emph{{Solvable models of quantum black holes:
  a review on Jackiw{\textendash}Teitelboim gravity}},
  \href{https://doi.org/10.1007/s41114-023-00046-1}{\emph{Living Rev. Rel.}
  {\bfseries 26} (2023) 4} [\href{https://arxiv.org/abs/2210.10846}{{\ttfamily
  2210.10846}}].

\bibitem{Erdos:2014zgc}
L.~Erd{\H{o}}s and D.~Schr{\"o}der, \emph{{Phase Transition in the Density of
  States of Quantum Spin Glasses}},
  \href{https://doi.org/10.1007/s11040-014-9164-3}{\emph{Math. Phys. Anal.
  Geom.} {\bfseries 17} (2014) 441}
  [\href{https://arxiv.org/abs/1407.1552}{{\ttfamily 1407.1552}}].

\bibitem{Cotler:2016fpe}
J.~S. Cotler, G.~Gur-Ari, M.~Hanada, J.~Polchinski, P.~Saad, S.~H. Shenker
  et~al., \emph{{Black Holes and Random Matrices}},
  \href{https://doi.org/10.1007/JHEP05(2017)118}{\emph{JHEP} {\bfseries 05}
  (2017) 118} [\href{https://arxiv.org/abs/1611.04650}{{\ttfamily
  1611.04650}}].

\bibitem{Berkooz:2018qkz}
M.~Berkooz, P.~Narayan and J.~Simon, \emph{{Chord Diagrams, Exact Correlators
  in Spin Glasses and Black Hole Bulk Reconstruction}},
  \href{https://doi.org/10.1007/JHEP08(2018)192}{\emph{JHEP} {\bfseries 08}
  (2018) 192} [\href{https://arxiv.org/abs/1806.04380}{{\ttfamily
  1806.04380}}].

\bibitem{Lin:2022rbf}
H.~W. Lin, \emph{{The Bulk Hilbert Space of Double Scaled SYK}},
  \href{https://doi.org/10.1007/JHEP11(2022)060}{\emph{JHEP} {\bfseries 11}
  (2022) 060} [\href{https://arxiv.org/abs/2208.07032}{{\ttfamily
  2208.07032}}].

\bibitem{Lin:2023trc}
H.~W. Lin and D.~Stanford, \emph{{A Symmetry Algebra in Double-Scaled Syk}},
  \href{https://doi.org/10.21468/SciPostPhys.15.6.234}{\emph{SciPost Phys.}
  {\bfseries 15} (2023) 234}
  [\href{https://arxiv.org/abs/2307.15725}{{\ttfamily 2307.15725}}].

\bibitem{Blommaert:2023opb}
A.~Blommaert, T.~G. Mertens and S.~Yao, \emph{{Dynamical Actions and
  Q-Representation Theory for Double-Scaled Syk}},
  \href{https://doi.org/10.1007/JHEP02(2024)067}{\emph{JHEP} {\bfseries 02}
  (2024) 067} [\href{https://arxiv.org/abs/2306.00941}{{\ttfamily
  2306.00941}}].

\bibitem{Blommaert:2024ymv}
A.~Blommaert, T.~G. Mertens and J.~Papalini, \emph{{The Dilaton Gravity
  Hologram of Double-Scaled Syk}},
  \href{https://doi.org/10.1007/JHEP06(2025)050}{\emph{JHEP} {\bfseries 06}
  (2025) 050} [\href{https://arxiv.org/abs/2404.03535}{{\ttfamily
  2404.03535}}].

\bibitem{Blommaert:2024whf}
A.~Blommaert, A.~Levine, T.~G. Mertens, J.~Papalini and K.~Parmentier,
  \emph{{An Entropic Puzzle in Periodic Dilaton Gravity and Dssyk}},
  \href{https://doi.org/10.1007/JHEP07(2025)093}{\emph{JHEP} {\bfseries 07}
  (2025) 093} [\href{https://arxiv.org/abs/2411.16922}{{\ttfamily
  2411.16922}}].

\bibitem{Heller:2024ldz}
M.~P. Heller, J.~Papalini and T.~Schuhmann, \emph{{Krylov Spread Complexity as
  Holographic Complexity Beyond Jackiw-Teitelboim Gravity}},
  \href{https://doi.org/10.1103/spcr-jgm6}{\emph{Phys. Rev. Lett.} {\bfseries
  135} (2025) 151602} [\href{https://arxiv.org/abs/2412.17785}{{\ttfamily
  2412.17785}}].

\bibitem{Heller:2025ddj}
M.~P. Heller, F.~Ori, J.~Papalini, T.~Schuhmann and M.-T. Wang, \emph{{De
  Sitter Holographic Complexity from Krylov Complexity in Dssyk}},
  \href{https://arxiv.org/abs/2510.13986}{{\ttfamily 2510.13986}}.

\bibitem{Bossi:2024ffa}
L.~Bossi, L.~Griguolo, J.~Papalini, L.~Russo and D.~Seminara,
  \emph{{Sine-Dilaton Gravity Vs Double-Scaled Syk: Exploring One-Loop Quantum
  Corrections}}, \href{https://doi.org/10.1007/JHEP06(2025)152}{\emph{JHEP}
  {\bfseries 06} (2025) 152}
  [\href{https://arxiv.org/abs/2411.15957}{{\ttfamily 2411.15957}}].

\bibitem{Alfinito:2026cky}
E.~Alfinito and M.~Beccaria, \emph{{Higher-Loop Wormhole Length in Sine-Dilaton
  Gravity from Dssyk Krylov Complexity}},
  \href{https://arxiv.org/abs/2606.20220}{{\ttfamily 2606.20220}}.

\bibitem{Fu:2025kkh}
Y.~Fu, H.-S. Jeong, K.-Y. Kim and J.~F. Pedraza, \emph{{Toward Krylov-Based
  Holography in Double-Scaled Syk}},
  \href{https://doi.org/10.1007/JHEP05(2026)056}{\emph{JHEP} {\bfseries 05}
  (2026) 056} [\href{https://arxiv.org/abs/2510.22658}{{\ttfamily
  2510.22658}}].

\bibitem{Okuyama:2023iwu}
K.~Okuyama, \emph{{High Temperature Expansion of Double Scaled Syk}},
  \href{https://doi.org/10.1016/j.physletb.2023.138036}{\emph{Phys. Lett. B}
  {\bfseries 843} (2023) 138036}
  [\href{https://arxiv.org/abs/2304.01522}{{\ttfamily 2304.01522}}].

\bibitem{Verg2013}
M.~J. Vergès, \emph{Cumulants of the q-semicircular law, tutte polynomials,
  and heaps}, \href{https://doi.org/10.4153/cjm-2012-042-9}{\emph{Canadian
  Journal of Mathematics} {\bfseries 65} (2013) 863–878}.

\bibitem{Goel:2023svz}
A.~Goel, V.~Narovlansky and H.~Verlinde, \emph{{Semiclassical Geometry in
  Double-Scaled Syk}},
  \href{https://doi.org/10.1007/JHEP11(2023)093}{\emph{JHEP} {\bfseries 11}
  (2023) 093} [\href{https://arxiv.org/abs/2301.05732}{{\ttfamily
  2301.05732}}].

\bibitem{Okuyama:2023bch}
K.~Okuyama and K.~Suzuki, \emph{{Correlators of Double Scaled Syk at
  One-Loop}}, \href{https://doi.org/10.1007/JHEP05(2023)117}{\emph{JHEP}
  {\bfseries 05} (2023) 117}
  [\href{https://arxiv.org/abs/2303.07552}{{\ttfamily 2303.07552}}].

\bibitem{Okuyama:2025fhi}
K.~Okuyama, \emph{{Non-Perturbative Corrections in the Semi-Classical Limit of
  Double-Scaled Syk}},
  \href{https://doi.org/10.1007/JHEP06(2025)044}{\emph{JHEP} {\bfseries 06}
  (2025) 044} [\href{https://arxiv.org/abs/2501.15501}{{\ttfamily
  2501.15501}}].

\bibitem{Xu:2024gfm}
J.~Xu, \emph{{On Chord Dynamics and Complexity Growth in Double-Scaled SYK}},
  \href{https://doi.org/10.1007/JHEP06(2025)259}{\emph{JHEP} {\bfseries 06}
  (2025) 259} [\href{https://arxiv.org/abs/2411.04251}{{\ttfamily
  2411.04251}}].

\bibitem{Berkooz:2024ifu}
M.~Berkooz, R.~Frumkin, O.~Mamroud and J.~Seitz, \emph{{Twisted Times, the
  Schwarzian and Its Deformations in Dssyk}},
  \href{https://doi.org/10.1007/JHEP05(2025)080}{\emph{JHEP} {\bfseries 05}
  (2025) 080} [\href{https://arxiv.org/abs/2412.14238}{{\ttfamily
  2412.14238}}].

\bibitem{Beccaria:2024szi}
M.~Beccaria and A.~Cabo-Bizet, \emph{{Giant Graviton Expansion of Schur Index
  and Quasimodular Forms}},  \href{https://arxiv.org/abs/2403.06509}{{\ttfamily
  2403.06509}}.

\bibitem{Billo:2013jba}
M.~Billo, M.~Frau, L.~Gallot, A.~Lerda and I.~Pesando, \emph{{Modular anomaly
  equation, heat kernel and S-duality in $N=2$ theories}},
  \href{https://doi.org/10.1007/JHEP11(2013)123}{\emph{JHEP} {\bfseries 11}
  (2013) 123} [\href{https://arxiv.org/abs/1307.6648}{{\ttfamily 1307.6648}}].

\bibitem{Bajc:2025jjv}
B.~Bajc and K.~Trailovi{\'c}, \emph{{Holographic thermal propagator from
  modularity}}, \href{https://doi.org/10.1007/JHEP11(2025)133}{\emph{JHEP}
  {\bfseries 11} (2025) 133}
  [\href{https://arxiv.org/abs/2509.02226}{{\ttfamily 2509.02226}}].

\bibitem{Stanford:2017thb}
D.~Stanford and E.~Witten, \emph{{Fermionic Localization of the Schwarzian
  Theory}}, \href{https://doi.org/10.1007/JHEP10(2017)008}{\emph{JHEP}
  {\bfseries 10} (2017) 008}
  [\href{https://arxiv.org/abs/1703.04612}{{\ttfamily 1703.04612}}].

\end{thebibliography}\endgroup
\bibliographystyle{JHEP-v2.9}
\end{document}